\documentclass[10pt,twocolumn,twoside]{IEEEtran}

\usepackage{lipsum}
\usepackage{amsmath,amsfonts,amssymb}
\usepackage{amsmath}
\usepackage{algorithmic}
\usepackage{algorithm}
\usepackage{array}
\usepackage{stfloats}
\usepackage{url}
\usepackage{verbatim}
\usepackage{graphicx}
\usepackage{cite}
\usepackage{color}
\usepackage{xcolor}
\usepackage{tikz}

\usetikzlibrary{arrows.meta, positioning}

\usepackage[capitalize]{cleveref}

\def\R{\mathbb{R}}
\def\Z{\mathbb{Z}}
\def\ugen{u^{\text{gen}}}
\def\Sc{\mathcal{S}}
\def\EE{\mathbf{E}}
\def\Gc{\mathcal{G}}
\newtheorem{proposition}{Proposition}
\newtheorem{remark}{Remark}
\newtheorem{theorem}{Theorem}
\newtheorem{definition}{Definition}
\newtheorem{corollary}{Corollary}
\newtheorem{lemma}{Lemma}

\DeclareMathOperator{\Equaldef}{\overset{def}{=}}

\usepackage[caption=false,font=footnotesize]{subfig}

\begin{document}

\title{Multi-Agent Coordination under Poisson Observations: A Global Game Approach}

\author{Marcos M. Vasconcelos and Behrouz Touri
\thanks{M. M. Vasconcelos is with the Department of Electrical and Computer Engineering, FAMU-FSU College of Engineering, Florida State University, Tallahassee, FL -- USA. E-mail: \texttt{m.vasconcelos@fsu.edu}. 

B. Touri is with the Department of Industrial Systems Engineering, University of Illinois, Urbana-Champaign, IL -- USA. E-mail: \texttt{touri1@illinois.edu}.}}

\maketitle

\begin{abstract}

We study a model of strategic coordination based on a class of games with incomplete information known as \textit{Global Games}. Under the assumption of Poisson-distributed signals and a Gamma prior distribution on state of the system, we demonstrate the existence of a Bayesian Nash equilibrium within the class of threshold policies for utility functions that are linear in the agents' actions. Although computing the exact threshold that constitutes an equilibrium in a system with finitely many agents is a highly non-trivial task, the problem becomes tractable by analyzing the game's potential function with countably infinitely many agents. Through numerical examples, we provide evidence that the resulting potential function is unimodal, exhibiting a well-defined maximum. Our results are applicable to the modeling of bacterial Quorum Sensing systems, whose noisy observation signals are often well-approximated using Poisson processes. 
\end{abstract}


\section{Introduction}

Across both artificial and biological distributed systems, the agents are often required to make task-oriented decisions based on noisy, incomplete information about their environments. In many such settings, the successful execution of a task requires the alignment of decisions — a phenomenon known as \textit{coordinated behavior}. While observation noise is ubiquitous, its specific probabilistic model can vary significantly depending on the application. In engineering, the Gaussian channel is the most common and analytically tractable model of noisy observation. However, many real-world applications are not accurately represented by the Gaussian model. For instance, in bacterial decision-making, cells can only sense discrete molecular signals which are better modeled as Poisson arrival processes. Similarly, in artificial systems that rely on optical sensors, photodetectors integrate discrete photon events, which are also naturally represented by Poisson random variables. Therefore, the development and analysis of multi-agent systems that must coordinate their decisions under Poisson observations is an important problem with a wide range of applications.

In this paper, we address the strategic coordination of agents equipped with sensors that measure discrete quantities modeled as Poisson random variables. Our framework is based on \textit{Global Games}, a class of stochastic coordination games with partial information, where the agents observe the state of the system through noisy channels and make binary decisions whether to make a \textit{risky} or \textit{costly} decision or not  \cite{Morris:2003}. Global Games with Gaussian channels and Gaussian prior distribution on the state are well-studied. However, very little is known about Global Games with different probabilistic models such as in the case of Poisson channels. The present collection of results aim at taking the first steps into  providing a theoretical basis to non-Gaussian models for strategic stochastic coordination. As a byproduct, our work can be used to model the distributed bacterial decision-making mechanism known as Quorum Sensing, which is ubiquitous in cellular decision making and is an integral component of many engineered biological systems exhibiting collective behavior \cite{Boedicker:2015}.

\subsection{Global Games -- Stochastic Coordination Games}

Coordination games have been widely studied by economists and engineers alike to model and predict outcomes in strategic settings when rational agents have incentive to align their decisions \cite{Dahleh:2016,Mahdavifar:2017,Paarporn:2021,Paarporn:2021b,Arditti:2024}. Among the class of coordination games, the payoff relevant terms (e.g. the state) may be perfectly observed or not. When the state is observed through stochastic observation channels, we obtain a Global Game (GG). The analysis of GGs show that a threshold strategy profile survives the process of iterated deletion of strictly dominated strategies, thus forming a Bayesian Nash Equilibrium (BNE) \cite{Carlsson:1993,Morris:2003,Chamley:2004,Fudenberg:1998}.  In engineering, GGs have been applied to model distributed task allocation in multi-robot systems \cite{Kanakia:2016,Vasconcelos:2023,Beaver:2025} and medium-access control in cognitive radio networks \cite{Krishnamurthy:2009}. More broadly, GGs are particularly well-suited for applications involving coordination — i.e., decision alignment among agents in choosing between risky and safe actions, where outcomes depend on incomplete observation of the state of the world.

The base setting for a GG consists of a finite or infinite set of agents, making noisy observations of a common payoff relevant term (also known as \textit{fundamental} in the economics literature). Each agent uses a policy to choose between a safe or risky action such as to optimize its expected utility satisfying a property called \textit{strategic complementarity}. The existing analysis of GGs available in the literature relies on Gaussian observations and a Gaussian prior distribution on the fundamental. In the many different variants of the base model threshold policies on the observation play a paramount role. The goal is to establish the emergence of a BNE within the class of threshold strategies. However, the analysis of GGs under different probabilistic models, in particular the model with Poisson observations and Gamma prior distributions (Gamma-Poisson) considered herein, is very limited until now. 

The Gamma-Poisson model has been used in a centralized setting in \cite{Vasconcelos:2018}, in which an approach based on a \textit{social-planner} approach, where the policies used by every agent in the system are jointly optimized. From a decentralized system perspective, the social-planner approach does not validate nor justifies the emergence of an equilibrium threshold strategy. 

The first attempt to use a GG model with Poisson signals was made in \cite{Vasconcelos:2022}, assuming a degenerate uniform prior distribution on the state. Although this type of probabilistic model is widely assumed in economics and leads to a tractable analysis, its applicability in stochastic systems is limited because it does not allow us to compute fundamental quantities, such as the probability of an agent taking a particular action — an essential quantity for establishing the coordination efficiency of the underlying mechanism. The present paper focuses on establishing the existence of equilibria with a threshold structure, and the topic of coordination efficiency in GGs is deferred to future work. We provide a complete analysis of a decentralized coordination system based on a GG formulation under the general Gamma-Poisson model, with an application in systems biology, as it is a widely used model for molecular communications and cellular decision-making \cite{Fang:2021,Fang:2023,Shaska:2023}.

\subsection{Contributions}

The main contributions of the paper are summarized as follows:

\begin{enumerate}

\item We introduce the class of Global Games with a Gamma prior distribution on the state and Poisson observation channels. We characterize the structure of the best response policy, and show that threshold policies emerge as Bayesian Nash equilibria.

\item We establish a sufficient condition for which the equilibria exists. To that end, we show that an equivalent deterministic game where the agents select thresholds as actions is a potential game. Under the sufficient condition, the potential game is finite, which guarantees the existence of a Nash equilibrium of an equivalent deterministic game, and therefore, the existence of the Bayesian Nash equilibrium for the original stochastic game.

\item Since the problem of computing the equilibrium using best-response dynamics is computationally intractable for systems with a large number of agents, we use the symmetry of the problem to compute a\textbf{} Nash equilibrium in the limit when the number of agents is infinite, by computing the maximizing the mean-field potential function. Closed form expressions for such potential function can be derived but are numerically unstable. However, sample average approximations are easy to implement and allows us to compute near-optimal thresholds for many cases of interest.

\end{enumerate}

\subsection{Notation}

We adopt the following notation. Random variables are denoted by upper case letters such as $X$, and their realizations are denoted using lower case letters such as $x$. The probability of an event $\mathfrak{E}$ is denoted by $\mathbf{P}(\mathfrak{E})$, and the expected value of a random variable $X$ is denoted by $\mathbf{E}[X]$. The probability density function of a continuous random variable $X$ is denoted by $f_X$. The probability mass function for a discrete random variable $Y$ taking values in a countable set $\mathbb{Y}$, is denoted by $\mathbf{P}(Y=y),\ y\in{\mathbb{Y}}$. Functionals are denoted by calligraphic letters such as $\mathcal{F}$. A policy for the $i$-th agent is denoted using indexed Greek letters such as $\mu_i$. A collection of policies is called a policy profile and is represented by $\mu\Equaldef(\mu_1,\ldots,\mu_N)$. Moreover, 
\begin{align}
\mu_{-i}\Equaldef(\mu_1,\ldots,\mu_{i-1},\mu_{i+1},\ldots,\mu_N)
\end{align}
denotes the policy profile used by all the opponents of the $i$-th agent. Throughout this work, we are dealing frequently with binary vectors $a\in \{0,1\}^N$. For such vectors, we let $|a|=\sum_{i=1}^Na_i$.

\subsection{Organization}
The rest of the paper is organized as follows. In Section II, we introduce our model Global Games with Gamma prior and Poisson observations (GGGP), where we show that the deterministic game in the absence of randomness is a coordination game. In Section III, we define the Best-Response policy and analyze its structural properties. In Section IV, we prove the existence of a BNE within the class of threshold strategies. In Section V, we compute the mean-field potential function. Section VI shows a few examples that demonstrate that our theoretical results lead to stable numerical methods to compute the optimal threshold for systems with very large number of agents. In Section VII we discuss the application for GGGPs in bacterial Quorum Sensing. The paper concludes in Section VIII with open problems and future research directions.

\section{System Model}

Consider a system with $N$ agents, and denote the collection of all agents by $[N]\Equaldef \{1,2,\ldots,N\}$. We assume that $N$ can be arbitrarily large, however, unlike other models with infinitely many agents (e.g. \textit{population games} \cite{Arcak:2021}), our model always has a countable number of agents.

Each agent can take a binary action $a_i\in\{0,1\}$, which represents the decision to engage in a free or costly behavior, respectively. Let $a_i=0$ denote the $i$-th agent's decision to \textit{not activate}, and $a_i=1$ denote its decision to \textit{activate}.
The decision to activate or not leads to a \textit{utility}. 
We adopt the convention that the utility of not activating is normalized to zero, whereas the utility of activating depends on the number of agents who decide to activate and on the state variable  $x$. Herein, the structure of the agent $i$'s utility is of the following form
\begin{align}\label{eq:generalgame}
    \ugen_i(a_i,a_{-i},x)= a_i \cdot  g\big(|a|,x\big), \ \ i\in[N],
\end{align}
where $g:\Z^+\times \R\to \R$ is an arbitrary function.
A relatively general example of such utility functions, is the class of separable utility functions of the form 
\begin{multline}\label{eq:payoff}
    u_i(a_i,a_{-i},x)\Equaldef a_i \cdot \Bigg( b\Big(\sum_{j\in [N]}a_j \Big)-c(x)\Bigg)\\=a_i \cdot \Big( b\big(|a|\big)-c(x)\Big).
\end{multline}
Notice that the utility function exhibits a separable structure: the first term, referred to as the \textit{benefit}, is a nonnegative function dependent on the number of agents deciding to activate, while the second term represents the activation cost.

We assume the benefit function $b: \mathbb{Z} \rightarrow \mathbb{R}$ is a strictly increasing function.  
The activation cost is defined by a function $c: \mathbb{R}\rightarrow \mathbb{R}$ of the state variable $x$.
Our model can be used for applications where it is more advantageous for agents to activate when
$x$ is either large or small.
For example, if the cost $c(x)$ is decreasing in $x$, it is more advantageous to activate when $x$ is large.
In this paper, we restrict our analysis to cost functions of the power-law form, i.e.,
\begin{equation}\label{eq:power_law}
c(x) \Equaldef x^p, \quad p \in \mathbb{Z}.
\end{equation}
This class of functions is broad enough to capture many cases of interest. The dichotomy arises by allowing 
$p$ to be either positive or negative.

The utility function in our model exhibits a property known as \textit{strategic complementarity} \cite{Hoffman:2019}, meaning that the utility is strictly increasing in the number of agents that activate for all $x$ and strictly decreasing in $x$ for a fixed number of activating agents  regardless of $p$. 

The following analysis relies on the notion of a \textit{potential game} \cite{monderer1996potential}.

\vspace{5pt}

\begin{definition}[Potential Game]\label{lem:exact_potential}
Let $\mathcal{A}_i$ denote the action set of the $i$-th agent in a game with utility functions $u_i(a_i,a_{-i},x)$, $i\in[N]$. Let $\mathcal{A} = \mathcal{A}_1\times \cdots \times \mathcal{A}_N$. A game is an \textit{exact potential game} if there exists a \textit{potential function} {$\Phi$}: 
 {$\mathcal{A}\times \mathbb{R}\rightarrow \mathbb{R}$} such that 
\begin{multline}\label{eq:exact_potential}
    u_{i}(a'_{i},a_{-i},x)- u_{i}(a''_{i},a_{-i},x) \\ = \Phi(a'_{i},a_{-i},x)- \Phi(a''_{i},a_{-i},x),
\end{multline}
for all $x\in \mathbb{R}$,  $a_{i}',a_{i}''\in \mathcal{A}_{i}$, $a_{-i}\in \mathcal{A}_{-i}$,  $i \in [N]$.
\end{definition}

\subsection{Omniscient agents}\label{sec:omniscient}

Ideally, we want to design policies that mitigate the detrimental effects of observation noise. Therefore, we aim to take the correct action when the state 
$x$ is above or below an appropriate threshold. However, since the state is not directly observed, some efficiency loss is inevitable. The preliminary question considered herein is \textit{What is the optimal decision if every agent could perfectly observe the state variable 
$x$, i.e., if every agent were omniscient?}

Let us assume that the state $x$ is available to every agent. In this case, for any given and known $x$ to all agents the game is deterministic. Our first result shows that, for any given $x$, \cref{eq:generalgame} is a potential game. 

\vspace{5pt}

\begin{proposition}\label{prop:potential}
    For any fixed $x\in \R$, the $N$ player binary action game with the utility functions in  \cref{eq:generalgame} is a potential game for any arbitrary $g:\Z^+\times \R\to \R$. Consequently, for any $x\in \R$, the game with the payoff structure in \cref{eq:payoff} is a potential game. 
\end{proposition}

\vspace{5pt}

\begin{IEEEproof}
    Note that for a fixed $x$, a game with utility functions \cref{eq:generalgame} is a congestion game with two resources $0,1$ and action sets ${\mathcal{A}_i=\{\{0\},\{1\}\}}$ for all agents $i\in [N]$, and the congestion functions $c_0(x)=0$ for resource $0$ and $c_1(x)=g(x)$ for resource $1$. As a result, the game is a  potential game~\cite{monderer1996potential} with the potential function $\Phi:\{0,1\}^N\to \R$ given by
    \begin{align}\label{eqn:potentialgeneral}
        \Phi(a)=\sum_{i=1}^{|a|}g(i,x), \qquad a\not=\{0,\ldots,0\},
    \end{align} 
    and $\Phi(0,\ldots,0)=0$.
\end{IEEEproof}

Next we identify the set of Nash equilibria of the games with the general utility functions of the form in \cref{eq:generalgame} when we have an increasing function $g(\ell,x)$ of $\ell$. Note that if $g(\ell,x)$ is increasing in $\ell$ (for any fixed $x$), then the strategy profile $(0,\ldots,0)$ is a Nash equilibrium iff 
\[g(1,x)=u_1(1,0,\ldots,0)\leq u_1(0,0,\ldots,0)=0.\]
Similarly, $(1,1,\ldots,1)$ is a Nash equilibrium iff \[0=u_1(0,1,\ldots,1)\leq u_1(1,\ldots,1)=g(N,x).\] 
If $g(\ell,x)$ is an increasing function of $\ell$, therefore, at least one of the two profiles would be a Nash equilibrium. The following result shows that these are the only possible (pure) Nash equilibria of such general games. 
\begin{proposition}\label{prop:pureeq}
    For the general coordination game characterized by \cref{eq:generalgame}, with a strictly increasing function $g(\ell,x)$ of $\ell$ for any fixed $x$, the set of Nash equilibria of the game $\Sc_x$ satisfies
    \begin{equation}\label{eqn:Sx}
        \mathcal{S}_x\subseteq \big\{(0,\ldots,0),(1,\ldots,1)\big\}.
    \end{equation}
\end{proposition}
\begin{IEEEproof}
Suppose that $\tilde{a}$ is a Nash equilibrium. If $0<|\tilde{a}|<N$, then there are two agents $i,j$ with $\tilde{a}_i=0$ while $\tilde{a}_j=1$. Since $\tilde{a}$ is a Nash equilibrium, agent $i$ has no strict incentive to deviate and hence, 
    \begin{align}
        0&=\ugen_{i}(\tilde{a}_i=0,\tilde{a}_{-i},x)\geq \ugen_{i}(1,\tilde{a}_{-i},x)\cr 
        &=g(|\tilde{a}|+1,x)>g(|\tilde{a}|,x)=u_{j}(\tilde{a},x).
    \end{align}
    Therefore, agent $j$ is better off with action $0$, which is a contradiction. As a result, at an equilibrium we need to have either $|\tilde{a}|=0$ or $|\tilde{a}|=N$.
\end{IEEEproof}
As a consequence of the above result, for the coordination game characterized by utility functions of the form in \cref{eq:payoff}, if $b(\ell)$ is an increasing function, then \cref{eqn:Sx} holds.
Since there are at most two Nash equilibria in pure strategies, we consider the following policy for omniscient agents for our coordination game described by \cref{eq:payoff}, which always achieves the equilibrium with the best possible payoff for the agents.

\vspace{5pt}

\begin{definition}[Omniscient policy]
Let $x$ denote the state, and $N$ denote the total number of agents in the system. The omniscient policy is
\begin{equation}
a^{\star}_i(x) = \begin{cases}
1 & \ \ \text{if} \ \  g(N,x)>0 \\
0 & \ \ \text{otherwise.} 
\end{cases}
\end{equation}
\end{definition}

\vspace{5pt}

\begin{figure}[t!]
    \centering

\begin{tikzpicture}[scale=0.625] 

    \begin{scope}[xshift=0cm]
        \draw[->] (-0.5, 0) -- (5.5, 0) node[right] {$x$};
        \draw[->] (0, -0.5) -- (0, 5.5) node[above] {};

        \node[below left] at (0,0) {0};

        \def\const{2} 
        \def\increasing{\x*\x/5} 

        \pgfmathsetmacro{\intersect}{sqrt(10)}

        \fill[gray!20] (0, \const) -- (\intersect, \const) -- (\intersect, 0) -- (0, 0) -- cycle;

        \draw[thick, violet] (0, \const) -- (5, \const) node[above] {$b(N)$};

        \draw[thick, teal, domain=0:5, samples=100] 
            plot (\x, {\increasing}) node[above] {$c(x)$};

        \node at (1.35, 1.5) {$a_i^\star(x) = 1$}; 
    \end{scope}

    \begin{scope}[xshift=6.5cm]
        \draw[->] (-0.5, 0) -- (5.5, 0) node[right] {$x$};
        \draw[->] (0, -0.5) -- (0, 5.5) node[above] {};

        \node[below left] at (0,0) {0};

        \def\const{2} 
        \def\decreasing{1/(\x)} 

        \pgfmathsetmacro{\intersect}{(1/\const)}

        \fill[gray!20] (\intersect, 0) -- (5, 0) -- (5, \const) -- (\intersect, \const) -- cycle;

        \draw[thick, violet] (0, \const) -- (5, \const) node[above] {$b(N)$};

        \draw[thick, teal, domain=0.2:5, samples=100] 
            plot (\x, {\decreasing}) node[above] {$c(x)$};

        \node at (2.75, 1.25) {$a_i^\star(x) = 1$}; 
    \end{scope}

\end{tikzpicture}

\caption{Omniscient policies for agents with perfect observation of the state variable \( x \): When the cost is increasing in the colony density (\( p > 0 \)), the policy is of the low-threshold type (left); when the cost is decreasing in $x$ (\( p < 0 \)), the policy is of the high-threshold type (right).
}
    \label{fig:omniscient}
\end{figure}
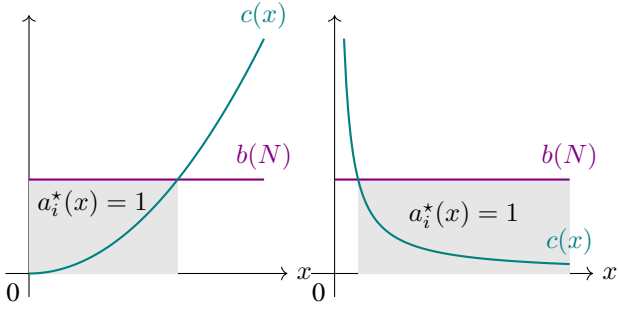

\vspace{5pt}

\begin{remark}
The omniscient policies obtained as a result of \cref{prop:pureeq} inform what we can expect from the analysis of the game with imperfect observations. More specifically, for 
utilities functions with the structure of \cref{eq:payoff}, we have
\begin{equation}
a^{\star}_i(x) = \begin{cases}
1 & \ \ \text{if} \ \  b(N) \geq c(x) \\
0 & \ \ \text{otherwise.} 
\end{cases}
\end{equation}
Therefore, for an increasing activation cost 
$c(x)$, the equilibrium policies are of the low-threshold type, while for a decreasing cost, the equilibrium policies are of the high-threshold type. This is illustrated in \cref{fig:omniscient}.
\end{remark}

\subsection{The state variable and its prior distribution}

The state $X$ which defines the activation cost is assumed to be a Gamma random variable. The Gamma distribution is the conjugate prior of the Poisson distribution, and is a natural probabilistic model for continuous quantities such as aggregate waiting times, intensities, concentrations, arrival rates, etc. With its flexibility and multiple degrees of freedom, the Gamma distribution allows for the fitting of shape and rate hyperparameters, making it suitable for modeling a wide range of probabilistic distributions over non-negative real numbers. 

Let $X$ be a Gamma random variable of shape $k > 0$, and rate $\theta>0$, that is
\begin{equation}
X \sim \mathcal{G}(k,\theta).
\end{equation}
 The probability density of a Gamma random variable is given by\footnote{For simplicity, we will assume that $k\in\mathbb{N}$. In this case, $$\Gamma(k)=(k-1)!$$} \begin{equation}
f_X(x) = \begin{cases} \frac{\theta^k}{\Gamma(k)}\cdot x^{k-1}\cdot e^{-\theta x} & x \geq 0 \\ 0 & \text{otherwise.}  \end{cases}
\end{equation}

\vspace{5pt}

\begin{remark}
Other prior distributions on the positive real numbers are also possible, such as
the folded Gaussian distribution. However, they lead to  non-tractable analysis. The Exponential distribution is particular case of the Gamma distribution when $k=1$.
\end{remark}

\subsection{Observation signals}

Once realized, the state is partially observed by each agent via independent Poisson channels. Given $X=x$, a Poisson arrival process of rate $\lambda x$ is generated, leading to the following random observation variable $Y_i \mid X=x$ at the $i$-th agent
\begin{align}
Y_i \mid X=x \sim \mathcal{P}(\lambda x), \ \ i\in[N].
\end{align}
Therefore,
\begin{align}
\mathbf{P}(Y_i=y\mid X=x) = \frac{(\lambda x)^y}{y!}e^{-\lambda x}, \ \ y \in \mathbb{Z}_{\geq 0}.
\end{align}

\vspace{5pt}

\begin{remark}
We assume that the observations $\{Y_i\}_{i\in[N]}$ are conditionally independent given $X=x$. We observe the following interesting features which are appropriate for this observation model: For a fixed state realization $X=x$, the variance of the signal increases with the rate $\lambda$. Therefore, a large $\lambda$ may degrade the accuracy of the system, instead of reducing uncertainty.
\end{remark}

\subsection{Activation policies and the optimization problem}

The last element in specifying our stochastic game formulation are the policies. Each agent acts solely on the basis of its observation $Y_i$. In our model, we assume that the agents do not share their observations with other agents (for examples, neighboring agents in a graph). More sophisticated Global Game models with information sharing exist \cite{Mahdavifar:2017,Dahleh:2016}, but their analysis is centered on the Gaussian case. The extension of our model with Poisson observations to an information sharing model is left for future work. 

The $i$-th agent's action $a_i$ is determined by a function 
$\mu_i:\mathbb{Z}_{\geq 0 } \rightarrow \{0,1\}$ of its private signal, i.e., 
\begin{equation}
    a_i= \mu_i(y), \ \ i \in[N].
\end{equation}
Let us define $\mathcal{M}_i$ as the set of all admissible policies for the $i$-th agent. The goal of each agent is to maximize its expected utility function with respect to its policy $\mu_i \in \mathcal{M}_i$, i.e.,
\begin{equation}\label{eq:expected_utility}
    \mathcal{J}_i(\mu_i,\mu_{-i}) \Equaldef \mathbf{E}\Big[ u_i\big(\mu_i(Y_i),\big\{\mu_j(Y_j)\big\}_{j\neq i},X\big) \Big].
\end{equation} 
Therefore, given the policies of other agents $\mu_{-i}$, agent $i$ strives to achieve
\begin{align}
\sup_{\mu_i \in \mathcal{M}_i} \ \ \mathcal{J}_i(\mu_i,\mu_{-i}).
\end{align}

In a system of self-motivated strategic decision-making agents in a stochastic setting, one of the solution concepts that correspond to this \textit{optimal behavior} is the notion of a Bayesian Nash-Equilibrium (BNE) \cite{Harsanyi:1967}.

\vspace{5pt}

\begin{definition}[Bayesian Nash-Equilibrium]
A strategy (policy) profile $\mu^\star$ is a Bayesian Nash-Equilibrium if
\begin{equation}
 \mathcal{J}_i(\mu^\star_i,\mu^\star_{-i}) \geq  \mathcal{J}_i(\mu_i,\mu^\star_{-i}), \  \ \mu_i\in\mathcal{M}_i,\ i\in[N],
\end{equation}
where $\mathcal{J}_i$ is the expected utility of the $i$-th agent defined in \cref{eq:expected_utility}.
\end{definition}

\section{Best response policies and their structure}

We begin by defining the class of threshold policies.

\vspace{5pt}

\begin{definition}[Threshold policies]\label{def:threshold}
A policy for the $i$-th agent is a threshold policy parameterized by $\tau_i \in \mathbb{Z}_{\geq 0}$ if it has one of the following forms:
\begin{equation}
    \mu^{\mathrm{low}}_i(y) \Equaldef \begin{cases}
    1 & \text{if} \ \ y\leq \tau_i\\
    0 & \text{otherwise}
    \end{cases}
\ \ \text{or} \ \ 
    \mu^{\mathrm{high}}_i(y) \Equaldef \begin{cases}
    1 & \text{if} \ \ y > \tau_i \\
    0 & \text{otherwise.}
    \end{cases}
\end{equation}

We express the threshold policy in a more compact form by using a unitary indicator function as $\mu^{\mathrm{low}}_i(y) = \mathbf{1}(y\leq \tau_i)$ and $\mu^{\mathrm{high}}_i(y) = \mathbf{1}(y> \tau_i).$
\end{definition}

\vspace{5pt}

\subsection{Best-response policies} 

Our analysis involves arguments based on the best-response to a given strategy profile. In particular, we are interested in establishing that within the class of threshold policy profiles there exists a BNE.

First, for an arbitrary agent $i\in[N]$, we define the best-response to an arbitrary policy profile $\mu_{-i}$. 
For an arbitrarily fixed strategy profile $\mu_{-i}$, we define the \textit{best-response} of the $i$-th agent to $\mu_{-i}$ as the function  $\mathcal{B}_{i}^{\mu_{-i}}:\mathbb{Z}_{\geq 0}\rightarrow \{0,1\}$ such that 
\begin{equation}
    \mathcal{B}_{i}^{\mu_{-i}}(y)  \Equaldef \arg \max_{\xi\in\{0,1\}}  \mathbf{E}\bigg[u_i\Big(\xi,\big\{\mu_j(Y_j)\big\}_{j\neq i}, X\Big)\ \Big| \ Y_i=y \bigg].
\end{equation}
Therefore, for the utility function in \cref{eq:payoff}, the best-response function is given by
\begin{equation}\label{eq:best_response}
  \mathcal{B}_{i}^{\mu_{-i}}(y)=
  \begin{cases}1&
    \mbox{if }\begin{aligned}[t]
       &\mathbf{E}\bigg[ b\Big( \sum_{j\neq i} \mu_j(Y_j) + 1 \Big) \ \Big| \   Y_i=y \bigg] \\
       &\qquad\qquad{\geq} \ \mathbf{E}\big[c(X)\mid Y_i=y\big],
       \end{aligned}\\
    0&\text{otherwise}.
  \end{cases}
\end{equation}

\vspace{5pt}
Notice that due to the separable structure of the utility function, there are two steps in the decision to activate or not when the $i$-th agent follows the best-response policy:

\begin{enumerate}

\item Estimating the activation cost, i.e., computing:

\begin{equation}
\hat{c}_i(y)\Equaldef \mathbf{E}\big[c(X)\mid Y_i=y\big].
\end{equation}

\item Estimating the benefit, i.e., computing:

\begin{equation}
\hat{b}_i(y)\Equaldef \mathbf{E}\bigg[ b\Big( \sum_{j\neq i} \mu_j(Y_j) + 1 \Big) \ \Big| \   Y_i=y \bigg].
\end{equation}
\end{enumerate}

\vspace{5pt}

The properties of these functions determine the structure of the best-response function. 
In particular, if the benefit function crosses the cost function exactly once as $y$ varies from  $0$ to  $\infty$ 
the best-response function exhibits a threshold structure, as the ones in \cref{def:threshold}.

\vspace{5pt}

\subsection{Activation Cost Estimate}

We proceed to characterize the estimate for the activation cost $c(X)$ given the agent's observed signal $Y=y$. The first step is to determine the conditional distribution of $X \mid Y_i=y$. 

\vspace{5pt}

\begin{lemma}\label{lem:conditional_distribution} Let $X\sim \mathcal{G}(k,\theta)$, and $Y_i\mid X=x \sim \mathcal{P}(\lambda x)$. Then, the posterior distribution of $X$ given $Y_i=y$ is 
\begin{equation}\label{eqn:prior}
X\mid Y_i=y \sim \mathcal{G}(y+k,\lambda+\theta).
\end{equation}
Moreover, the $i$-th agent's observation $Y_i$ is a Negative Binomial random variable with the following probability mass function:
\begin{equation}\label{eq:observed_signal1}
\mathbf{P}(Y_i=y) = \binom{y+k-1}{y}\cdot \left(\frac{{\theta}}{\lambda + {\theta}} \right)^k \cdot \left(\frac{\lambda}{\lambda+{\theta}} \right)^y,
\end{equation}
where $y\in\mathbb{Z}_{\geq 0}$.
\end{lemma}

\vspace{5pt}

\begin{IEEEproof}
The proof of \cref{eq:observed_signal1} follows from the conjugacy of Gamma distribution with regard to Poisson likelihood (see e.g. Section~3.2.3. in~\cite{robert2007bayesian}). The assertion about the posterior distribution of $Y_i$ follows from the fact that Poisson-Gamma mixture is negative Bionomial (see e.g., Section~8.2.1 in \cite{hilbe2011negative}).
\end{IEEEproof}

\vspace{5pt}

\begin{figure}[t!]
    \centering
    \includegraphics[width=0.8\columnwidth]{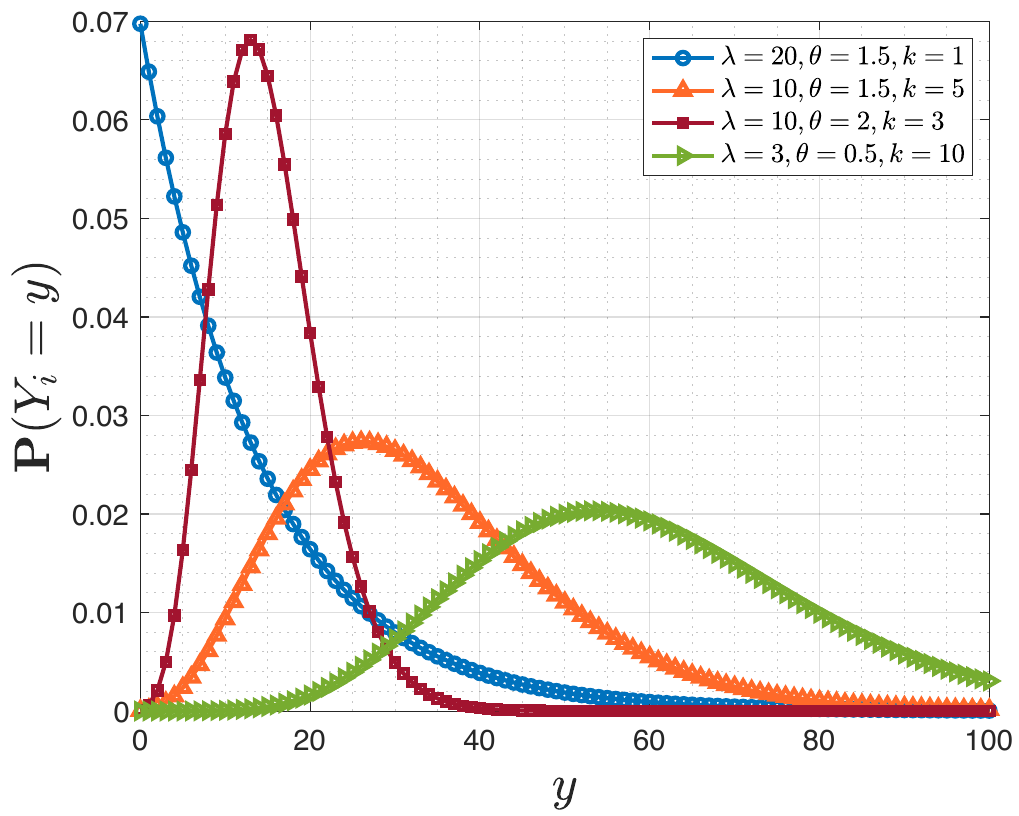}
    \caption{Examples of probability distribution functions for the number of AHL molecules observed by the $i$-th agent in the Gamma-Poisson Global Game model. The random variable $Y_i$ has a Negative Binomial distribution.}
    \label{fig:QS_Signal}
\end{figure}

\vspace{5pt}

We proceed by considering activation cost functions of the power-law type in \cref{eq:power_law}.
For such cost functions, the optimal estimates of the activation cost can be obtained in closed form. 

\vspace{5pt}

\begin{lemma}
Let $X\sim \mathcal{G}(k,\theta)$, and $Y\mid X=x \sim \mathcal{P}(\lambda x)$. If $c(x)=x^p$, then the $i$-th agent's optimal estimate for the activation cost given $Y_i=y$ is given by
\begin{equation}\label{eq:MMSE}
\hat{c}_i(y) = \frac{\Gamma(p+y+k)}{\Gamma(y+k)(\lambda+\theta)^p}, \ \ i\in[N],
\end{equation}
where \( \Gamma(z) \) is the Gamma function, the continuous extension of the factorial function for complex and real number arguments (excluding negative integers and zero), defined as
\begin{equation}
\Gamma(z) \Equaldef \int_0^{\infty} t^{z-1} e^{-t} \ dt,
\end{equation}
where \( z \) is a complex number with a real part greater than zero.
\end{lemma}

\vspace{5pt}

\begin{IEEEproof}
Recall that  $\hat{c}_i(y)=\EE[X^p\mid Y_i=y]$ which is the $p$-th moment of $X\mid Y_i=y$. By Lemma~\ref{lem:conditional_distribution}, \[X\mid Y_i=y\sim\mathcal{G}(y+k,\lambda+\theta).\] Therefore, the claim follows from the fact that the $p$-th moment $m_p$ of a $Z\sim \Gc(\alpha,\lambda)$ is $m_p=\frac{\Gamma(\alpha+p)}{\Gamma(\alpha)\lambda^p}$.
\end{IEEEproof}

\vspace{5pt}

Two special cases of interest are:
\begin{equation}\label{eq:positive}
p=1: \ \ \hat{c}_i(y) = \frac{y+k}{\lambda + {\theta}}.
\end{equation}
and
\begin{equation}\label{eq:negative}
p=-1: \ \ \hat{c}_i(y) = \frac{\lambda+{\theta}}{y+k-1}.
\end{equation}

\begin{figure}
    \centering
    \includegraphics[width=0.8\columnwidth]{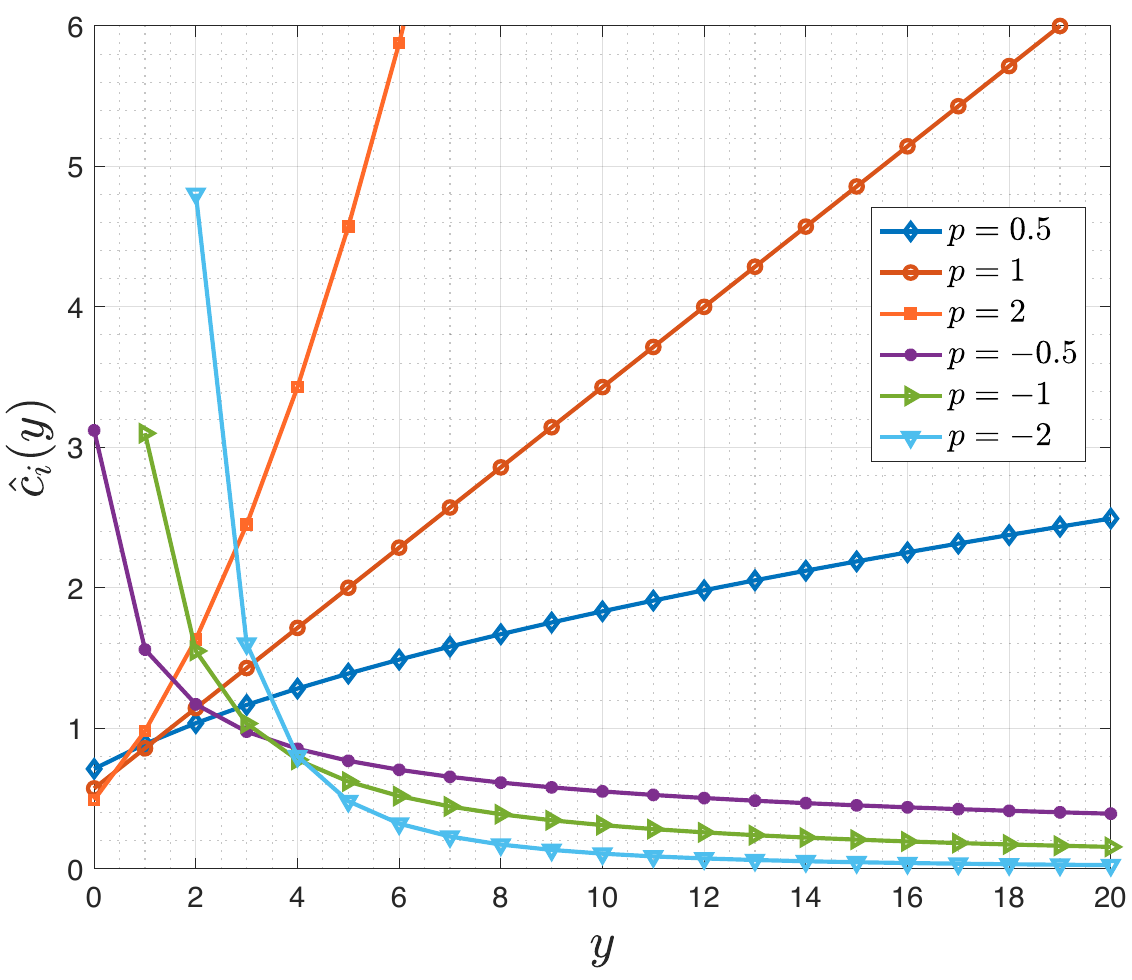}
    \caption{Monotonicity of the optimal activation cost estimate in the observed signal for different values of $p$ when $\lambda=3$,
${\theta} = 0.1$ and $
k = 1$.}
    \label{fig:Monotonicity}
\end{figure}

By inspecting \cref{eq:positive,eq:negative}, we observe that \( \hat{c}_i(y) \) satisfies an important property depending on the value of \( p \): 
if \( p = +1 \), the optimal activation cost estimate is strictly increasing in \( y \), and if \( p = -1 \), 
the optimal activation cost estimate is strictly  decreasing in \( y \).
 The next lemma formalizes and generalizes this property for any value of $p\in \mathbb{R}$. \Cref{fig:Monotonicity} illustrates the monotonicity of the optimal estimate functions for different values of $p$. 

\vspace{5pt}

\begin{lemma}\label{lem:increasing}
Let $k\in \mathbb{Z}_{\geq 1}$, ${\theta} \in \mathbb{R}_{>0}$, $\lambda \in \mathbb{R}_{>0}$ and $p\in \mathbb{R}$.  If $p>0$, the optimal activation cost estimate $\hat{c}_i(y)$ is strictly increasing for all $y \geq 0$. If $p<0$, the optimal activation cost estimate $\hat{c}_i(y)$ is strictly decreasing for $y \geq 1-p-k$.
\end{lemma}

\vspace{5pt}

\begin{IEEEproof}
Consider the first-order difference:
\begin{IEEEeqnarray}{rCl}
\Delta \hat{c}_i(y) & \Equaldef & \hat{c}_i(y+1) - \hat{c}_i(y) \nonumber \\
& = & \frac{1}{(\lambda+{\theta})^p} \Bigg[ \frac{\Gamma(p+y+1+k)}{(y+k)!} - \frac{\Gamma(p+y+k)}{(y+k-1)!} \Bigg] \nonumber \\
& \stackrel{(a)}{=} & \frac{p}{(\lambda+{\theta})^p} \cdot  \frac{\Gamma(p+y+k)}{\Gamma(y+k+1)}, 
\end{IEEEeqnarray}
where equality $(a)$ follows from Gamma function's \textit{fundamental property}: $\Gamma(z+1)=z\Gamma(z)$.

Note that if $p>0$, since $\Gamma(z)>0$ for all $z>0$,  $y\geq 0$, and $k\geq 1$, we have $\Delta \hat{c}_i(y) > 0$.

Similarly, for $p<0$, since $\Gamma(z)>0$ for all $z>0$, if $y\geq1-p-k$, we have $\Delta \hat{c}_i(y) < 0$.
\end{IEEEproof}

\subsection{Benefit Estimate}

The second quantity that needs to be estimated from the received signal $y$ by the $i$-th agent when determining its best response is the benefit, which depends on the actions of the remaining agents in the system. This step requires forming a belief on the action of any other agent $j\neq i$ given the measured signal. In this analysis, we assume that other agents are using a threshold policy characterized by a threshold ${\tau_j \in \mathbb{Z}_{\geq 0}}$. 

To establish the properties of the optimal benefit estimate, we must first define and characterize the notion of \textit{belief}. Upon observing $Y_i=y$, we define the belief of agent $i$ on the action of agent $j$, as
\begin{equation}
\pi_{ij}(y) \Equaldef \mathbf{P}(A_j=1 \mid Y_i=y).
\end{equation}

Since a threshold policy can be classified as ``low'' or ``high'', we define the beliefs accordingly:

\vspace{5pt}

\begin{equation}
\pi^{\mathrm{low}}_{ij}(y \mid \tau_j) \Equaldef \mathbf{P}(Y_j\leq \tau_j \mid Y_i=y).
\end{equation}

and

\begin{equation}
\pi^{\mathrm{high}}_{ij}(y \mid \tau_j) \Equaldef \mathbf{P}(Y_j > \tau_j \mid Y_i=y).
\end{equation}

\vspace{5pt}

\begin{lemma}[Belief on other agent's signals]\label{lem:identity}
Let ${\theta} \in \mathbb{R}_{>0}$, $\lambda \in \mathbb{R}_{>0}$, $k\in\mathbb{Z}_{\geq 1}$, and $y\in\mathbb{Z}$. For any other agent $j\neq i$, the belief on the observation $Y_j$ given $Y_i=y$ has the following Negative Binomial distribution
\begin{equation}
\mathbf{P}(Y_j=\ell \mid Y_i=y) = \binom{\ell+k+y}{\ell}  \Big(\frac{\lambda}{{\theta}+2\lambda} \Big)^\ell  \Big(\frac{{\theta}+\lambda}{{\theta}+2\lambda} \Big)^{k+y},
\end{equation}
for all $\ell \geq 0$.
\end{lemma}

\vspace{5pt}

\begin{IEEEproof}
The proof can be found in Appendix \ref{sec:belief}.
\end{IEEEproof}

\vspace{5pt}

As a direct consequence of \cref{lem:identity}, the belief on the action of an agent using a ``low'' threshold policy with parameter $\tau_j$ is
\begin{multline}\label{eq:belief_low}
    \pi_{ij}^{\mathrm{low}}(y \mid \tau_j)= \sum_{\ell=0}^{\tau_j} \binom{\ell+k+y-1}{\ell} \left(\frac{\lambda}{2\lambda+\theta}\right)^\ell \times \\ \left(\frac{\lambda +\theta}{2\lambda+\theta}\right)^{k+y}  .
\end{multline}

A similar expression can easily be derived for $\pi_{ij}^{\mathrm{high}}$ and since it can be readily obtained from $\pi_{ij}^{\mathrm{low}}$ it is omitted for brevity. From here on, we will assume that the benefit function has the following \textit{normalized linear} structure
\begin{equation}\label{eq:lin_ben}
b(\xi) = \frac{g}{N}\xi,
\end{equation}
where $g>0$ is the \textit{gain} constant, and $N$ is the number of agents. The interpretation of the normalized linear benefit is that what the agents produce as a result of their activation constitutes a \textit{public good}, whose benefit is shared among all the agents in the system \cite[and references therein]{Dandekar:2012,Heilmann:2015,Hota:2016}. For the rest of the paper, we constrain the analysis  to this case for simplicity of exposition. The extension to nonlinear benefit functions is tractable and left for future work.

Consider $i\in [N]$, and let all other agents $j\neq i$ use low threshold policies, $\mu_j^{\mathrm{low}}$. Then, for the normalized linear benefit function in \cref{eq:lin_ben}, we have
\begin{align}
\hat{b}_i(y)& =  \mathbf{E}\bigg[ \frac{g}{N}\Big( \sum_{j\neq i} \mu^{\mathrm{low}}_j(Y_j) + 1 \Big) \ \Big| \   Y_i=y \bigg]\\
 & =  \frac{g}{N} \left( \sum_{j\neq i} \pi^{\mathrm{low}}_{ij}(y\mid \tau_j) +1\right).  \label{eq:benefit_linear}
\end{align}
A similar expression holds when all other agents use $\mu_j^{\mathrm{high}}$ instead.
We proceed with showing that the functions $\hat{b}_i(y)$ are either strictly monotone increasing or decreasing depending on whether agents $j\neq i$ are all using a low or high threshold policy.

\vspace{5pt}

\begin{lemma}\label{lem:monotone_beliefs}
Let $\theta,\lambda \in \mathbb{R}_{>0}$, $k\in\mathbb{Z}_{\geq 1}$, $y\in\mathbb{Z}$ and $i\neq j$. The belief $\pi^{\mathrm{low}}_{ij}(y \mid \tau_j)$ given by \cref{eq:belief_low} is monotone decreasing in $y$ for all $\tau_j \in \mathbb{Z}_{\geq 0}$. Consequently, $\pi^{\mathrm{high}}_{ij}(y\mid \tau_j)$ is monotone increasing in $y$ for all $\tau_j \in \mathbb{Z}_{\geq 0}$.
\end{lemma}

\vspace{5pt}

\begin{IEEEproof}
The proof can be found in Appendix \ref{sec:monotonicity}.
\end{IEEEproof}

\vspace{5pt}

\begin{proposition}\label{prop:monotonicity}
Let $i\in [N]$. For normalized linear benefit functions, if all other agents $j\neq i$ are using low threshold policies, the optimal estimate of the benefit function $\hat{b}_i(y)$ is strictly monotone decreasing in $y$. If all other agents $j\neq i$ are using high threshold policies, the optimal estimate of the benefit function $\hat{b}_i(y)$ is strictly monotone increasing in $y$. Moreover, 
\begin{equation}
\hat{b}_i(\infty) = \begin{cases}
g/N, \  \text{if}  \ \mu_{-i} = (\mu_{1}^{\mathrm{low}}, \ldots, \mu_{i-1}^{\mathrm{low}},\mu_{i+1}^{\mathrm{low}},\ldots,\mu_{N}^{\mathrm{low}})\\
g, \  \text{if} \ \mu_{-i} = (\mu_{1}^{\mathrm{high}}, \ldots, \mu_{i-1}^{\mathrm{high}},\mu_{i+1}^{\mathrm{high}},\ldots,\mu_{N}^{\mathrm{high}}).
\end{cases}
\end{equation}
\end{proposition}

\vspace{5pt}

\begin{IEEEproof}
Recall that when all other agents in the system are using either low (or high) threshold functions, the optimal benefit estimate is given by \cref{eq:benefit_linear}
where $\pi^{\mathrm{low}}_{ij}(y)$ (or $\pi^{\mathrm{high}}_{ij}(y)$) is the belief that agent $j$ will activate, given that agent $i$ has observed a signal $Y_i=y$. 
From \cref{lem:monotone_beliefs}, the function $\hat{b}_i(y)$ is the nonnegative sum of strictly monotone decreasing (or increasing) functions, and therefore, it is strictly monotone decreasing (or increasing). 
\end{IEEEproof}

\vspace{5pt}

\section{Optimality of Threshold Policies}

Having established the monotonicity of the benefit and activation cost estimates, and their role in determining a best-response policy for an agent under the assumption that all other agents are following threshold strategies, we now state and prove the main result of the paper. In this section, we establish the 
\textit{single crossing property} \cite{Athey:2001} between benefit and activation cost estimates in the best-response policy when the benefit is a normalized linear function and the agents use threshold policies. Given this property, the best-response to any vector of thresholds is a threshold policy, and therefore the class of threshold policies are \textit{closed} under the best-response dynamics \cite{Swenson:2018}. The main consequence of this result is that a GGGP can be understood as a deterministic game where the agents are choosing thresholds \textit{ex-ante} from the set of non-negative integers as their actions \cite[and references therein]{Wu:2022}. Then, we obtain that under a sufficient condition on the parameters of the GGGP, we show that the best-response thresholds cannot be arbitrarily large, leading to an equivalent deterministic game with a finite action space. 

\vspace{5pt}

\begin{theorem}[Optimality of Low Threshold Strategies]
Consider the Global Game with $N>1$ agents, Poisson observations with parameter $\lambda$ and Gamma prior distribution with parameters $k,\theta$ on the state variable. Let $g>0$ such that $b(\xi)=(g/N)\cdot \xi$ and $c(x)=x^p$. If $p>0$, and 
\begin{equation}\label{eq:condition}
g > \frac{N}{(\lambda+\theta)^p} \cdot \frac{\Gamma(p+k)}{(k-1)!} \cdot \left[ (N-1)\cdot\left(\frac{\theta+\lambda}{\theta+2\lambda} \right)^k+1\right]^{-1},
\end{equation}
then there exists a BNE  strategy profile $\mu^\star$ where all the agents use a threshold policy, $\mu^{\mathrm{low}}$.   
\end{theorem}

\vspace{5pt}

\begin{IEEEproof}
For an arbitrary agent $i\in [N]$, fix a threshold strategy profile $\mu_{-i}^{\mathrm{low}} = (\mu_{1}^{\mathrm{low}}, \ldots, \mu_{i-1}^{\mathrm{low}},\mu_{i+1}^{\mathrm{low}},\ldots,\mu_{N}^{\mathrm{low}})$, where each agent other than $i$ is potentially using a different threshold, $\tau_j$, $j\neq i$. From \cref{prop:monotonicity}, the benefit estimate function $\hat{b}_i(y)$ is a strictly monotone decreasing function of $y$. If $p>0$, \cref{lem:increasing} implies that the activation cost estimate $\hat{c}_i(y)$ is strictly monotone increasing in $y$. Recall that the best-response is equal to 1 if $\hat{b}_i(y)\geq \hat{c}_i(y)$, and zero otherwise. Since $\hat{c}_i(y)$ is unbounded, if
\begin{equation}\label{eq:sufficient}
\hat{b}_i(0) > \hat{c}_i(0), 
\end{equation}
 the two functions will cross exactly a single time.

When $Y_i=0$, i.e., in the absence of Poisson arrivals, we have
\begin{equation}
\hat{b}_i(0) = \frac{g}{N} \cdot \left( \sum_{i\neq j} \pi^{\mathrm{low}}_{ij}(0 \mid \tau_j) +1\right),
\end{equation}
where
\begin{equation}\label{eq:belief_at_zero}
   \pi_{ij}^{\mathrm{low}}(0 \mid \tau_j)= \sum_{\ell=0}^{\tau_j} \binom{\ell+k-1}{\ell}  \left(\frac{\lambda}{2\lambda+\theta}\right)^\ell\left(\frac{\lambda +\theta}{2\lambda+\theta}\right)^{k}. 
\end{equation}
Since all the terms in \cref{eq:belief_at_zero} are positive, we can lower bound it as
\begin{equation}
   \pi_{ij}^{\mathrm{low}}(0 \mid \tau_j) \geq \pi_{ij}^{\mathrm{low}}(0 \mid 0) = \left(\frac{\lambda +\theta}{2\lambda+\theta}\right)^{k}.
\end{equation}
Therefore, we obtain the following nontrivial lower bound to the benefit function at zero
\begin{equation} \label{eq:bound_benefit}
\hat{b}_i(0) \geq \frac{g}{N} \cdot \left( (N-1) \left(\frac{\lambda +\theta}{2\lambda+\theta}\right)^{k} +1\right).
\end{equation}
Similarly, 
\begin{equation}
\hat{c}_i(0) = \frac{\Gamma(p+k)}{(k-1)!(\lambda+\theta)^p}.
\end{equation}

Let $\tau_i^{\star}$ be defined as:
\begin{equation}
\tau_i^{\star} = \max \big\{ y \mid \hat{b}_i(y) > \hat{c}_i(y)\big\}.
\end{equation}
From \cref{eq:bound_benefit}, we have that \cref{eq:condition} implies \cref{eq:sufficient}. Then, the threshold that characterizes the best-response policy to this strategy for agent $i$ must be finite.

Suppose that within the fixed threshold strategy profile $\mu_{-i}^{\mathrm{low}}$, there are infinite thresholds. Let $\mathbb{J}_{\infty}$ be the set of agents with such thresholds.  In this case, 
\begin{equation}
\hat{b}_i(y) =  \frac{g}{N} \left( \sum_{i\neq j, \\ \ i\notin \mathbb{J}_\infty} \pi^{\mathrm{low}}_{ij}(y) + |\mathbb{J}_{\infty}| +1\right),
\end{equation}
which is strictly monotone decreasing. Suppose $|\mathbb{J}_\infty|<N$, then
\begin{equation}
\lim_{y\rightarrow \infty} \hat{b}_i(y) = \frac{g}{N}(|\mathbb{J}_\infty|+1) \leq g.
\end{equation}

Since $\lim_{y\rightarrow \infty}\hat{c}_i(y)=\infty$, we must have
\begin{equation}
\tau_i^{\star} < \infty.
\end{equation}

We can repeat this argument for all agents in $\mathbb{J}_\infty$, and obtain a new strategy profile where every agent uses a finite threshold.
\end{IEEEproof}

\vspace{5pt}

\begin{corollary}\label{cor:finite}
There exists a sufficiently large integer \( \bar{T} \in \mathbb{Z} \) such that the best response to any finite collection of threshold strategies is itself a threshold strategy with values in the lattice \( \{0, 1, \dots, \bar{T}\} \).
\end{corollary}

\vspace{5pt}

\begin{IEEEproof}
The proof follows by recalling that for $p>0$, the function $\hat{c}_i(y)$
is strictly monotone increasing, and $\hat{b}_i(y)$
is strictly monotone decreasing in $y$ and strictly monotone increasing in each $\tau_j$ characterize $\mu_{-i}^{\mathrm{low}}$. Therefore, the
worst-case scenario is when $\tau_j=\infty$ for all $j\neq i$, in which case we have: $\hat{b}_i(y) = g$. Since $\hat{c}_i(y)$ is strictly increasing, there will exist a value $\bar{T}$ such that $\hat{c}_i(y) < g$ for all $y\leq \bar{T}$ and $\hat{c}_i(y) \geq g$ for all $y> \bar{T}$. Since $\hat{b}_i(y)\leq g$, we must have that $\tau_i^\star \leq \bar{T}$ for all $i\in [N]$. Therefore, as long as the sufficient condition in \cref{eq:condition} holds, $\bar{T}$ is an upper bound for $\tau^\star_i,$ for all $i\in [N]$. 
\end{IEEEproof}

\vspace{5pt}

Lastly, we address the case when $p<0$. The analysis is similar with a few minor changes in the argument. 

\vspace{5pt}

\begin{theorem}[Optimality of High Threshold Strategies]
Consider the Global Game with Poisson observations with parameter $\lambda$ and Gamma prior distribution with parameters $k,\theta$ on the state. Let $g>0$ such that $b(\xi)=(g/N)\cdot \xi$ and $c(x)=x^p$. If $p<0$, and 
\begin{equation}\label{eq:condition2}
g < \frac{(\lambda+\theta)^{-p}}{\Gamma(1-p)}\bigg[ 1- \bigg(\frac{N-1}{N}\bigg)(1-p)\Big(\frac{\theta+\lambda}{\theta+2\lambda} \Big)^{1-p}   \bigg]^{-1},
\end{equation}
then there exists a BNE  strategy profile $\mu^\star$ where all the agents use a threshold policy, $\mu^{\mathrm{high}}$.  
\end{theorem}

\vspace{5pt}

\begin{IEEEproof}
For an arbitrary agent $i\in [N]$, fix a threshold strategy profile $\mu_{-i}^{\mathrm{high}} = (\mu_{1}^{\mathrm{high}}, \ldots, \mu_{i-1}^{\mathrm{high}},\mu_{i+1}^{\mathrm{high}},\ldots,\mu_{N}^{\mathrm{high}})$, where each agent other than $i$ is potentially using a different threshold, $\tau_j$, $j\neq i$. From \cref{prop:monotonicity}, the benefit estimate function $\hat{b}_i(y)$ is a strictly monotone increasing function of $y$. If $p<0$, \cref{lem:increasing} implies that the activation cost estimate $\hat{c}_i(y)$ is strictly monotone decreasing in $y$. Recalling that the best-response is equal to 1 if $\hat{b}_i(y)\geq \hat{c}_i(y)$, and zero otherwise. If
\begin{equation}\label{eq:sufficient2}
\hat{b}_i(1-p-k) < \hat{c}_i(1-p-k), 
\end{equation}
since $\lim_{y\rightarrow \infty}\hat{c}_i(y)=0$, the two functions will cross exactly a single time. Let $\tau_i^{\star}$ be defined as
\begin{equation}
\tau_i^{\star} = \max \big\{ y \mid \hat{b}_i(y) < \hat{c}_i(y)\big\}.
\end{equation}

The following upper bound holds
\begin{equation}
\pi_{ij}^{\mathrm{high}}(y\mid \tau_j) \leq 1-(k+y)\bigg(\frac{\theta+\lambda}{\theta+2\lambda} \bigg)^{k+y},
\end{equation}
with equality if and only if $\tau_j=0$. From \cref{lem:increasing}, there is a critical value $y_{\mathrm{crit}}=1-p-k$ that determines whether the benefit estimate has exactly one crossing point with the cost estimate. Thus,
\begin{equation}
\pi_{ij}^{\mathrm{high}}(1-p-k\mid \tau_j) \leq 1-(1-p)\bigg(\frac{\theta+\lambda}{\theta+2\lambda} \bigg)^{1-p}.
\end{equation}

Therefore, we obtain the following nontrivial upper bound to the benefit function at $1-p-k$:

\begin{equation} \label{eq:upper_bound_benefit}
\hat{b}_i(1-p-k) \leq \frac{g}{N} \cdot \left( N-(N-1)(1-p)\bigg(\frac{\theta+\lambda}{\theta+2\lambda} \bigg)^{1-p}\right).
\end{equation}
Moreover,
\begin{equation}
\hat{c}_i(1-p-k\mid \tau_j) = \frac{1}{\Gamma(1-p)}\bigg(\frac{1}{\theta+\lambda} \bigg)^{p}.
\end{equation}

Suppose that within the fixed threshold strategy profile $\mu_{-i}^{\mathrm{high}}$, there are infinite thresholds. Let $\mathbb{J}_{\infty}$ be the set of agents other than $i$ with such thresholds. In this case, 
\begin{equation}
\hat{b}_i(y) =  \frac{g}{N} \left( \sum_{j\neq i, \\ \ j\notin \mathbb{J}_\infty} \pi^{\mathrm{high}}_{ij}(y\mid \tau_j) + 1 \right),
\end{equation}
which is strictly monotone increasing. Moreover, from \cref{eq:upper_bound_benefit} we have that \cref{eq:condition2} implies \cref{eq:sufficient2}, and that for any $|\mathbb{J}_\infty|<N$, we have
\begin{equation}
\lim_{y\rightarrow \infty} \hat{b}_i(y) = \frac{g}{N}(N-|\mathbb{J}_\infty|+1) > 0.
\end{equation}
Since $\lim_{y\rightarrow \infty}\hat{c}_i(y)=0$, we must have
\begin{equation}
\tau_i^{\star} < \infty.
\end{equation}
Therefore, the threshold that characterizes the best-response policy to this strategy for agent $i$ must be finite.

We can repeat this argument for all agents in $\mathbb{J}_\infty$, and obtain a new strategy profile where every agent uses a finite threshold.
\end{IEEEproof}

\section{Equivalent Deterministic Games}

The previous section established the existence of a BNE in the class of threshold policies. That is an important structural result because it allows the agents to select a threshold as if it were an action \textit{ex-ante} in an equivalent deterministic game \cite{Wu:2022,Harsanyi:1967,Harsanyi:1968}. We proceed with showing that this equivalent game is a finite \textit{exact potential game} and, as such, admits a Nash equilibrium in pure strategies \cite{monderer1996potential,Marden:2009,Marden:2012}. That is, there exists a set of thresholds for the entire system to which the agents will converge to if they use a best-response learning dynamics \cite{Fudenberg:1998}. We will focus on the case $p>0$ and $\mu^{\mathrm{low}}$ policies. The same arguments and results hold in the case of $p<0$ and $\mu^{\mathrm{high}}$ policies with minor modifications. 

Consider a deterministic game with $N>1$ agents. Assume that each agent picks a threshold $\tau_i \in \{0,1,\ldots, \bar{T}\}\Equaldef \mathcal{T}_i$, where $\bar{T}\in \mathbb{Z}$ is the upper bound on thresholds guaranteed to exist by \cref{cor:finite}. The $i$-th agent's utility is determined by 
\begin{multline}\label{eq:equivalent}
U_i(\tau_i,\tau_{-i}) \Equaldef  \mathbf{E}\Bigg[ \mathbf{1}(Y_i\leq \tau_i) \times \\ \Big(\frac{g}{N}\big(\sum_{j\neq i}\mathbf{1}(Y_j\leq \tau_j)+1\big) -c(X)\Big) \Bigg],
\end{multline}
where the expectation is over the joint probability distribution of $(X,Y_1,\ldots,Y_N)$ induced by the Poisson-Gamma model described in Section II.

\vspace{5pt}

\begin{lemma}\label{lem:potential}
Let $x \in \mathbb{R}$. Consider the deterministic binary coordination game indexed by $x$ with  agent set $[N]$, where the $i$-th agent has action space $\mathcal{A}_i = \{0,1\}$, and utility $u_i: \mathcal{A}_1 \times \cdots \times \mathcal{A}_N  \times \mathbb{R} \rightarrow \mathbb{R}$ defined as 
\begin{equation}
u_i(a_i,a_{-i},x)\Equaldef a_i \cdot \Bigg( \frac{g}{N}\cdot \sum_{j\in [N]}a_j -c(x)\Bigg), \ \ i\in[N].
\end{equation}
This game admits an exact potential function 
\begin{equation}\label{eq:potential1}
\Phi(a,x) = \frac{1}{2}\sum_{i=1}^N \sum_{j\neq i} \phi_{ij}(a_i,a_j,x),
\end{equation}
where
\begin{equation}\label{eq:potential2}
\phi_{ij}(a_i,a_{j},x) \Equaldef \frac{g}{N}a_ia_j + \Big(\frac{a_j+a_i-1}{N-1} \Big)\cdot \Big(\frac{g}{N}-c(x)\Big).
\end{equation}
\end{lemma}

\vspace{5pt}

\begin{IEEEproof}
This result can be verified using \cref{lem:exact_potential} and is omitted for brevity.
\end{IEEEproof}

\vspace{5pt}

\begin{theorem}
Let $\boldsymbol{\tau} \in \mathcal{T}=\mathcal{T}_1 \times \cdots \times \mathcal{T}_N$. The deterministic  game with utilities given by \cref{eq:equivalent} and action space $\mathcal{T}$ is an exact potential game with potential function
\begin{equation}\label{eq:potential3}
\tilde{\Phi}_N(\boldsymbol{\tau}) \Equaldef \mathbf{E} \Bigg[ \Phi \Big(\big\{\mathbf{1}(Y_i\leq \tau_i)\big\}_{i=1}^N,X \Big) \Bigg],
\end{equation}
where the expectation is over the joint probability distribution of $(X,Y_1,\ldots,Y_N)$ induced by the Poisson-Gamma model.
\end{theorem}

\vspace{5pt}

\begin{IEEEproof}
Let $\tau_i',\tau_i'' \in \mathcal{T}_i$. Consider
\begin{equation}
    \Delta(\tau_i',\tau_i'' \mid \tau_{-i}) \Equaldef U_i(\tau_i',\tau_{-i}) - U_i(\tau_i'',\tau_{-i}),
\end{equation}
which is equal to
\begin{multline}
\Delta(\tau_i',\tau_i'' \mid \tau_{-i}) = \mathbf{E}\Bigg[ \Big( \mathbf{1}(Y_i\leq \tau_i')-\mathbf{1}(Y_i\leq \tau_i'')\Big)\\ \Big(\frac{g}{N}\big(\sum_{j\neq i}\mathbf{1}(Y_j\leq \tau_j)+1\big) -c(X)\Big) \Bigg].
\end{multline}
However,
\begin{multline}
\Delta(\tau_i',\tau_i'' \mid \tau_{-i}) = \int_{0}^{\infty}\sum_{y\in\mathbb{Z}^N} \mathbf{P}(X=x,Y=y)\times \\ \Bigg[ \Big( \mathbf{1}(y_i\leq \tau_i')-\mathbf{1}(y_i\leq \tau_i'')\Big) \\ \Big(\frac{g}{N}\big(\sum_{j\neq i}\mathbf{1}(y_j\leq \tau_j)+1\big) -c(x)\Big) \Bigg] dx.
\end{multline}

Using \cref{lem:potential}, since for each possible value for the observations $\{y_i\}$, the resulting actions $\{\mathbf{1}(y_i\leq \tau_i')\}$ and $\{\mathbf{1}(y_i\leq \tau_i'')\}$ are binary, we obtain the following identity
\begin{multline}
\Delta(\tau_i',\tau_i'' \mid \tau_{-i}) = \int_{0}^{\infty}\sum_{y\in\mathbb{Z}^N} \mathbf{P}(X=x,Y=y)\times \\ \Big[ \Phi(\mathbf{1}(y_i\leq \tau_i'),\{\mathbf{1}(y_j\leq \tau_j)\}_{j\neq i},x) \\ - \Phi(\mathbf{1}(y_i\leq \tau_i''),\{\mathbf{1}(y_j\leq \tau_j)\}_{j\neq i},x) \Big]dx,
\end{multline}
which is equal to
\begin{multline}
\Delta(\tau_i',\tau_i'' \mid \tau_{-i}) = \mathbf{E}\Big[ \Phi(\mathbf{1}(Y_i\leq \tau_i'),\{\mathbf{1}(Y_j\leq \tau_j)\}_{j\neq i},X)\Big]  \\ - \mathbf{E}\Big[ \Phi(\mathbf{1}(Y_i\leq \tau_i''),\{\mathbf{1}(Y_j\leq \tau_j)\}_{j\neq i},X) \Big] \\ = \tilde{\Phi}(\tau_i',\tau_{-i})-\tilde{\Phi}(\tau_i'',\tau_{-i}).
\end{multline}

\end{IEEEproof}

\vspace{5pt}

\begin{corollary}
If \cref{eq:condition} is satisfied, the GGGP admits an equivalent finite exact potential game and, as a result, has a BNE in threshold policies.
\end{corollary}

\vspace{5pt}

\section{Global Games with Infinitely Many Agents}

The maximizers of the potential function of \cref{{eq:potential3}} are Nash Equilibria of the equivalent ex-ante deterministic game, where the agents choose thresholds as their actions \cite{monderer1996potential}. Therefore, to identify the BNE of a particular GGGP, we would like to solve the following optimization problem
\begin{equation}
\boldsymbol{\tau}^\star \in \arg \max_{\boldsymbol{\tau}\in \{0,\ldots,\bar{T}\}^N} \tilde{\Phi}_N(\boldsymbol{\tau}),
\end{equation}
However, solving this problem is difficult because of its discrete domain space that grows exponentially with the number of agents with a finite and unknown maximum threshold, $\bar{T}$. We circumvent this difficulty by analyzing the system with a large number of homogeneous agents. Such abstractions are particularly useful and appropriate in the study of swarms and bacterial colonies, where the number of agents is very large.

In this section, the structure of the potential function is explored to solve the optimization problem with countably infinitely many agents, i.e., $N\rightarrow \infty$ and obtain a Nash Equilibrium for the game \cite{Sanjari:2020,Mahajan:2013}.

\vspace{5pt}

\begin{definition}[Mean-Field Potential Function]
Assume a homogeneous system, where all the agents use the same threshold policy indexed by $\tau \in \mathbb{Z}_{\geq 0}$. Then, the \textit{mean-field potential function} (MFPF) is defined as
\begin{equation}
\Phi_{MF} (\tau) \Equaldef \lim_{N\rightarrow \infty} \frac{1}{N}\tilde{\Phi}_N\big((\tau,\ldots,\tau)\big).
\end{equation}

\end{definition}

\vspace{5pt}

\begin{theorem}
Let $X\sim \mathcal{G}(k,\theta)$, and $Y \mid X \sim \mathcal{P}(\lambda X)$, the mean-field potential function for the corresponding GGGP with a countably infinite number of agents is
\begin{multline}\label{eq:MFPF}
\Phi_{MF} (\tau) = \frac{g}{2}\mathbf{E}\Big[\mathbf{P}^2(Y\leq \tau \mid X) \Big]  \\ - \mathbf{E}\Bigg[\Bigg( \mathbf{P}(Y \leq \tau\mid X) -\frac{1}{2}\Bigg)\cdot c(X)  \Bigg],
\end{multline}
where the expectations are taken with respect to $X$.
\end{theorem}

\vspace{5pt}

\begin{IEEEproof}
Consider the potential function defined by \cref{eq:potential1,eq:potential2,eq:potential3} when every agent uses the same threshold $\tau$
\begin{multline}
\tilde{\Phi}_N (\tau) =  \frac{1}{2}\sum_{i=1}^N \sum_{j\neq i}\mathbf{E}\Bigg[ \frac{g}{N}\mathbf{1}(Y_i \leq \tau) \mathbf{1}(Y_j \leq \tau) \\ + \Big(\frac{\mathbf{1}(Y_j \leq \tau)+\mathbf{1}(Y_i \leq \tau)-1}{N-1}\Big)\cdot \Big(\frac{g}{N}-c(X)\Big)\Bigg],
\end{multline}
where the expectations are taken over $(Y_1,\ldots,Y_N,X)$. Using iterated expectations, we obtain
\begin{multline}
\tilde{\Phi}_N (\tau) = \frac{1}{2}\sum_{i=1}^N \sum_{j\neq i}\mathbf{E}\Bigg[ \mathbf{E}\bigg[ \frac{g}{N}\mathbf{1}(Y_i \leq \tau) \mathbf{1}(Y_j \leq \tau) \\ + \Big(\frac{\mathbf{1}(Y_j \leq \tau)+\mathbf{1}(Y_i \leq \tau)-1}{N-1}\Big)\cdot \Big(\frac{g}{N}-c(X)\Big) \ \bigg| \ X \bigg] \Bigg].
\end{multline}
From the conditional independence between $Y_i$ and $Y_j$ given $X=x$, we have
\begin{multline}
\tilde{\Phi}_N (\tau) =  \frac{1}{2}\sum_{i=1}^N \sum_{j\neq i}\mathbf{E}\Bigg[ \frac{g}{N}  \mathbf{P}(Y_i \leq \tau\mid X) \mathbf{P}(Y_j \leq \tau \mid X) \\ + \Big(\frac{\mathbf{P}(Y_j \leq \tau\mid X)+\mathbf{P}(Y_i \leq \tau\mid X)-1}{N-1}\Big)\cdot \Big(\frac{g}{N}-c(X)\Big)  \Bigg].
\end{multline}
Since $\{Y_i\mid X=x\}_{i=1}^N$ are identically distributed, we have
\begin{multline}
\tilde{\Phi}_N (\tau) = \frac{1}{2}\sum_{i=1}^N \mathbf{E}\Bigg[ \frac{g}{N}  \mathbf{P}(Y_i \leq \tau\mid X) \sum_{j\neq i} \mathbf{P}(Y_j \leq \tau \mid X) \\ + \Bigg( \frac{\sum_{j\neq i}\mathbf{P}(Y_j \leq \tau\mid X)}{N-1} + \mathbf{P}(Y_i \leq \tau\mid X)-1\Bigg)\\ \cdot \Big(\frac{g}{N}-c(X)\Big)  \Bigg],
\end{multline}
which leads to
\begin{multline}
\tilde{\Phi}_N (\tau) =  \frac{1}{2} \mathbf{E}\Bigg[ g (N-1)  \mathbf{P}^2(Y \leq \tau \mid X)  \\ + \Bigg( 2N \mathbf{P}(Y \leq \tau\mid X) -N\Bigg)\cdot \Big(\frac{g}{N}-c(X)\Big)  \Bigg],
\end{multline}
where $Y$ is a generic random variable which has the same distribution of $Y_i$ when conditioned on $X$, $i\in[N]$. Thus,
normalizing the resulting expression gives 
\begin{multline}
\frac{1}{N}\tilde{\Phi}_N(\tau) = \frac{1}{2} \mathbf{E}\Bigg[ g \frac{(N-1)}{N}  \mathbf{P}^2(Y \leq \tau \mid X)  \\ + \Bigg( 2 \mathbf{P}(Y \leq \tau\mid X) -1\Bigg)\cdot \Big(\frac{g}{N}-c(X)\Big)  \Bigg].
\end{multline}
Finally, taking the limit on $N$, we obtain
\begin{multline} 
\lim_{N\rightarrow \infty}\frac{1}{N}\tilde{\Phi}_N (\tau) =  \frac{g}{2} \mathbf{E}\Big[  \mathbf{P}^2(Y \leq \tau \mid X) \Big] \\ - \mathbf{E} \Bigg[\Bigg( \mathbf{P}(Y \leq \tau\mid X) -\frac{1}{2}\Bigg)\cdot c(X)  \Bigg].
\end{multline}

\end{IEEEproof}

\vspace{5pt}

\begin{remark}
Although, the mean-field potential function for the GGGP can be expressed in closed form, the resulting formula is not insightful and is computationally unstable, because it involves the evaluation of double factorial operators, frequently resulting in overflow errors. Fortunately, \cref{eq:MFPF} can be accurately estimated by sampling from the Gamma distribution of $X$ and computing the empirical mean. Additionally, the function $\Phi_{MF}$ over $\tau\in\{0,1,\ldots,\bar{T}\}$ is nonconcave, even when the domain is relaxed to be the interval $[0,\bar{T}]$ (c.f. \cref{sec:numerics}). The advantage of using \cref{eq:MFPF} to compute the optimal threshold is to avoid iterative procedures based on best-response dynamics, which could result in extremely long convergence time when the number of agents is very large.
\end{remark}

\subsection{Numerical Examples}\label{sec:numerics}

In this section we use the MFPF approach to compute the optimal threshold for a large population of agents. Recall that a sufficient condition for the existence result in Theorem 1 requires that \cref{eq:condition} is satisfied. When $N\rightarrow \infty$, \cref{eq:condition} becomes:
\begin{equation}
g>\frac{(\theta+2\lambda)^k}{(\theta+\lambda)^{p+k}}\cdot \frac{\Gamma(p+k)}{\Gamma(k)}.
\end{equation}

\begin{figure}
    \centering
    \includegraphics[width=0.8\columnwidth]{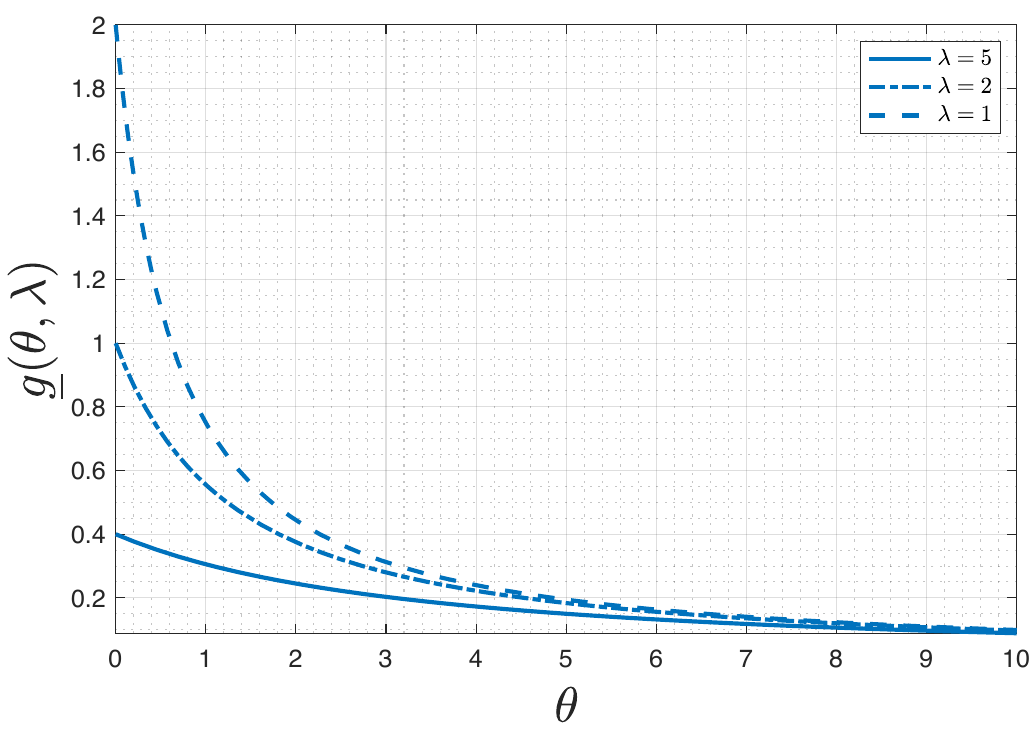}
    \caption{Critical value $\underline{g}(\theta,\lambda)$ for the existence of a BNE threshold, $\tau^\star$ when $p=k=1$ when $N\rightarrow \infty$.}
    \label{fig:Existence}
\end{figure}

One particular case of interest is obtained when $k=1$, i.e., the density of the colony is exponentially distributed, leading to signals $Y_i$ with a Geometric distribution. We assume that $p=1$, i.e., the activation cost grows linearly with the state. In that case, the sufficient condition simplifies to
\begin{equation}
g>\frac{\theta+2\lambda}{(\theta+\lambda)^{2}}\Equaldef \underline{g}(\theta,\lambda).
\end{equation}
\Cref{fig:Existence} displays the critical value $\bar{g}(\lambda,\theta)$ above which the gains are guaranteed to lead to existence of BNE threshold policies when $k=1$.

    \begin{figure*}[ht!]
    \centering
    \setkeys{Gin}{width=0.295\linewidth}
\subfloat[Potential Function for $k=1$, $\theta=1$, $\lambda=5$ and gain $g=1,2,3$]{\includegraphics{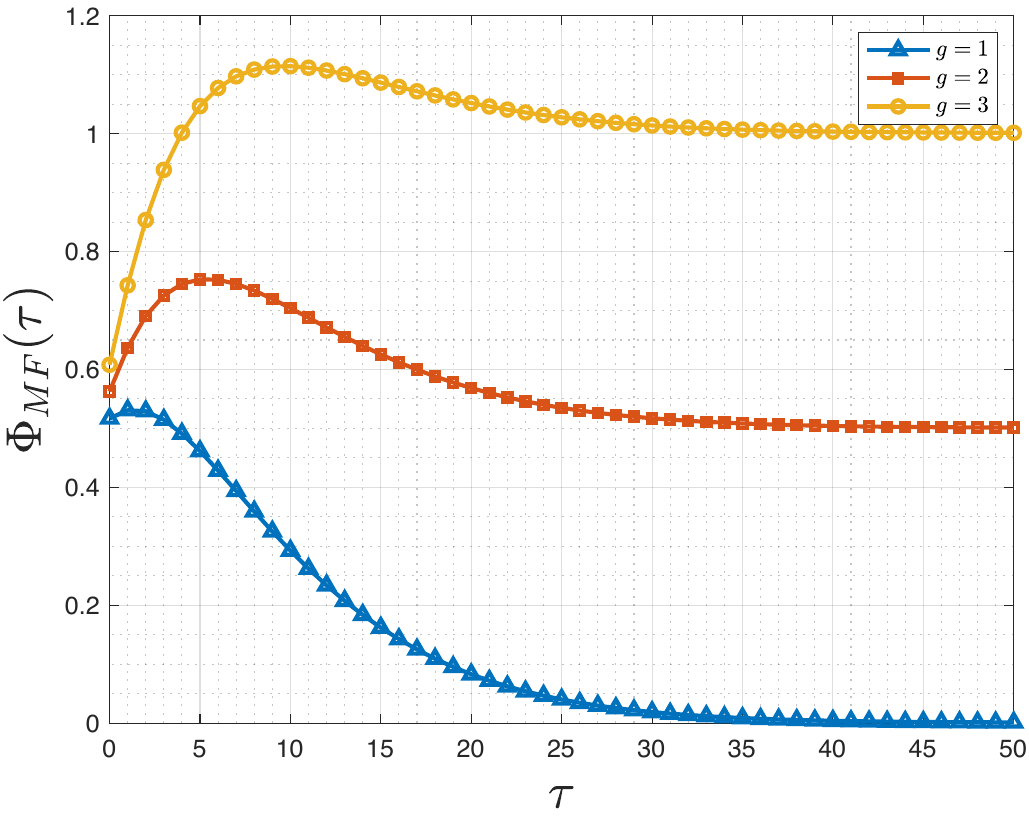}}
\label{MFPF1}
\hfill
\subfloat[Potential Function for $k=2$, $\theta=0.1$, $\lambda=5$ and gain $g=1,10,20$]{\includegraphics{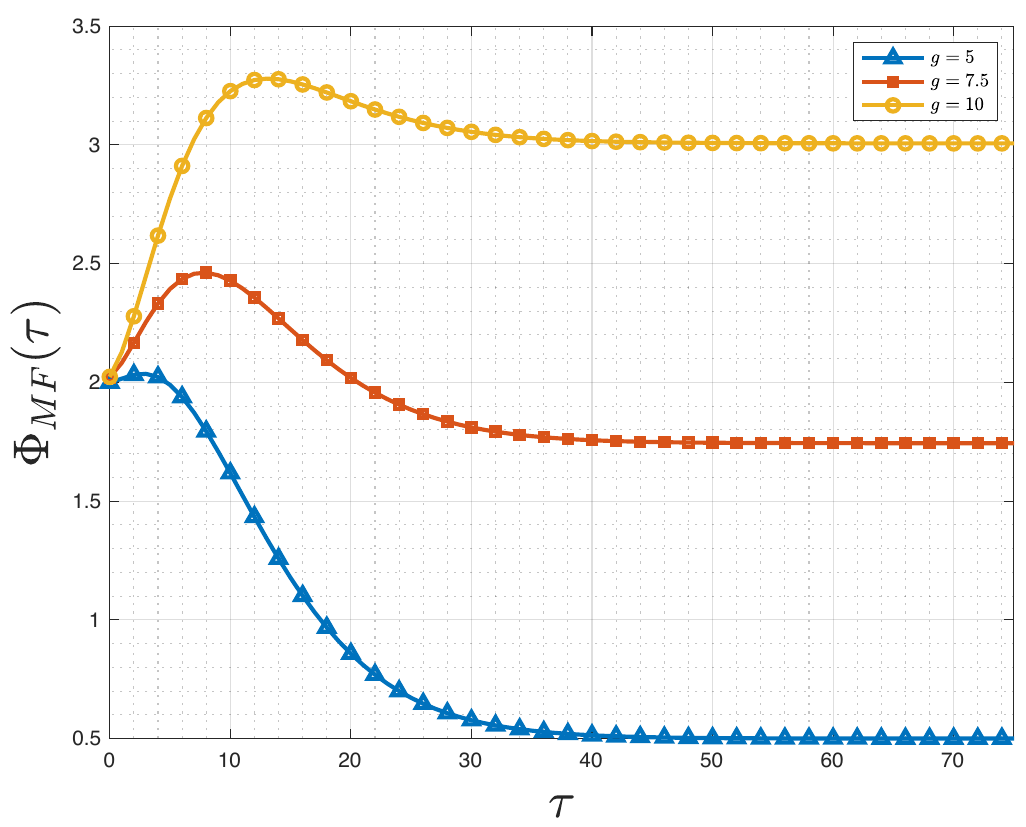}}
\hfill
\subfloat[Potential Function for $k=3$, $\theta=0.1$, $\lambda=5$ and gain $g=1,10,20$]{\includegraphics{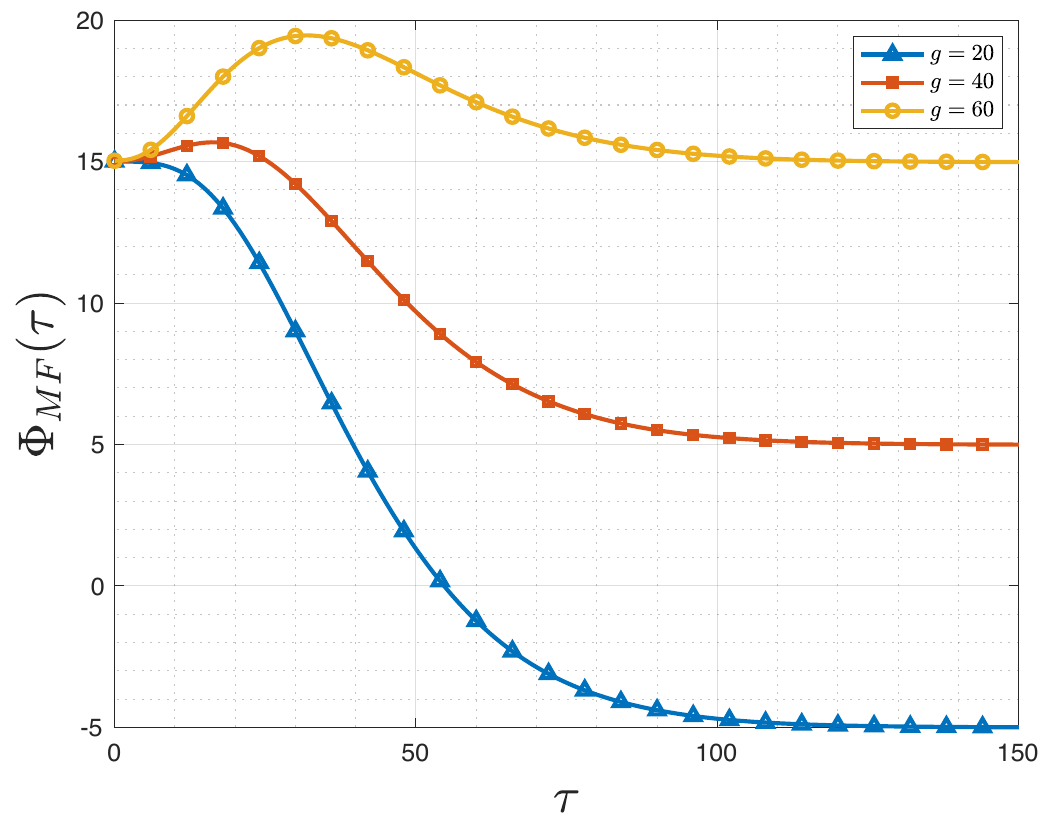}}
\hfill
\subfloat[Signal distribution for $k=1$, $\theta=1$, $\lambda=5$ and corresponding optimal thresholds.]{\includegraphics{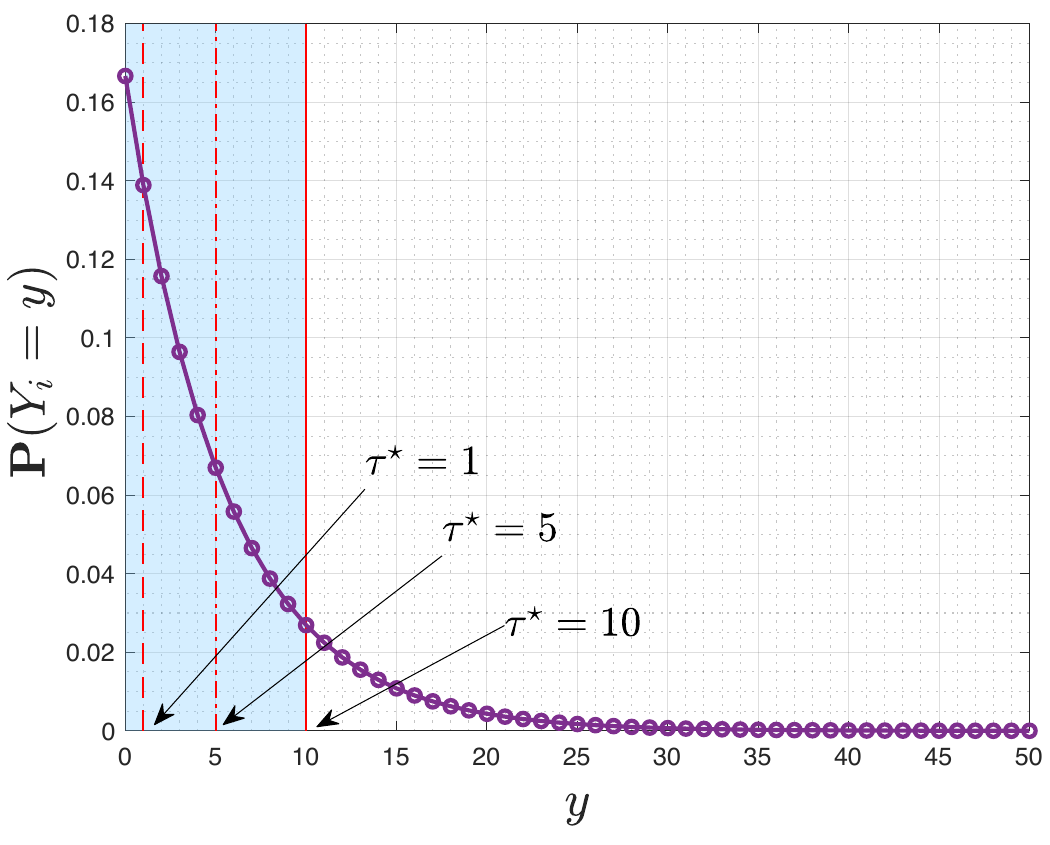}}
\hfill
\subfloat[Signal distribution for $k=2$, $\theta=0.1$, $\lambda=5$ and corresponding optimal thresholds.]{\includegraphics{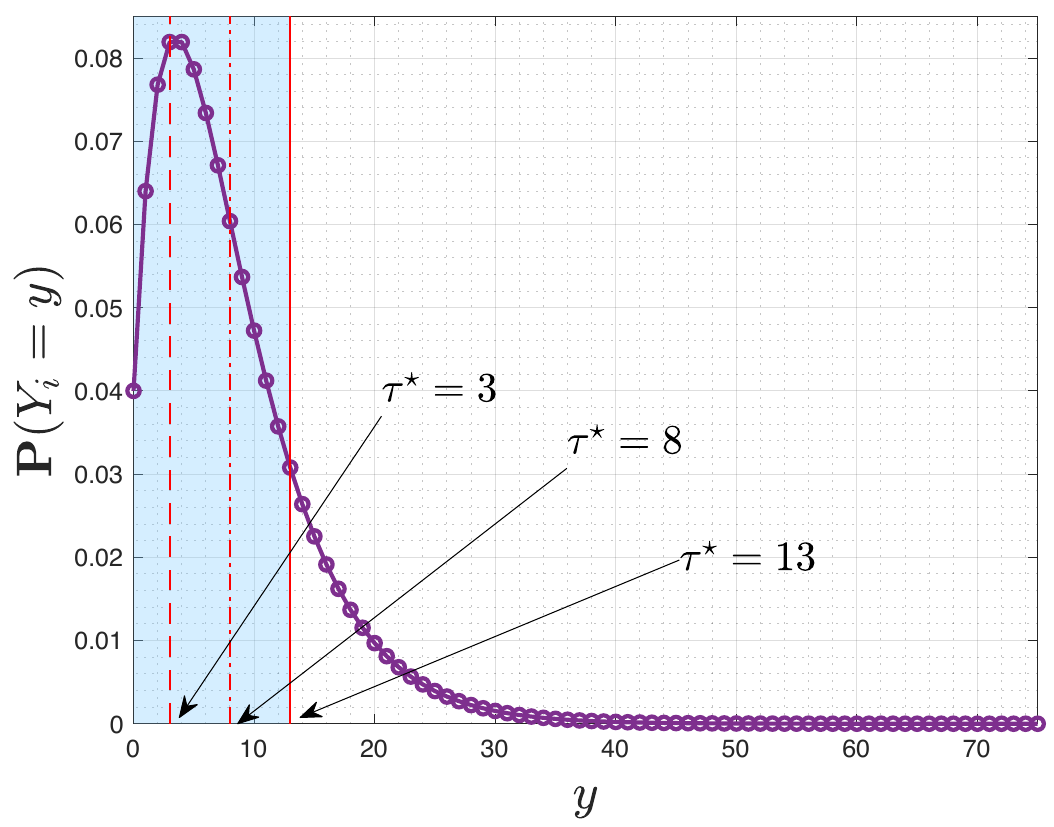}}
\hfill\subfloat[Signal distribution for $k=3$, $\theta=0.1$, $\lambda=5$ and corresponding optimal thresholds.]{\includegraphics{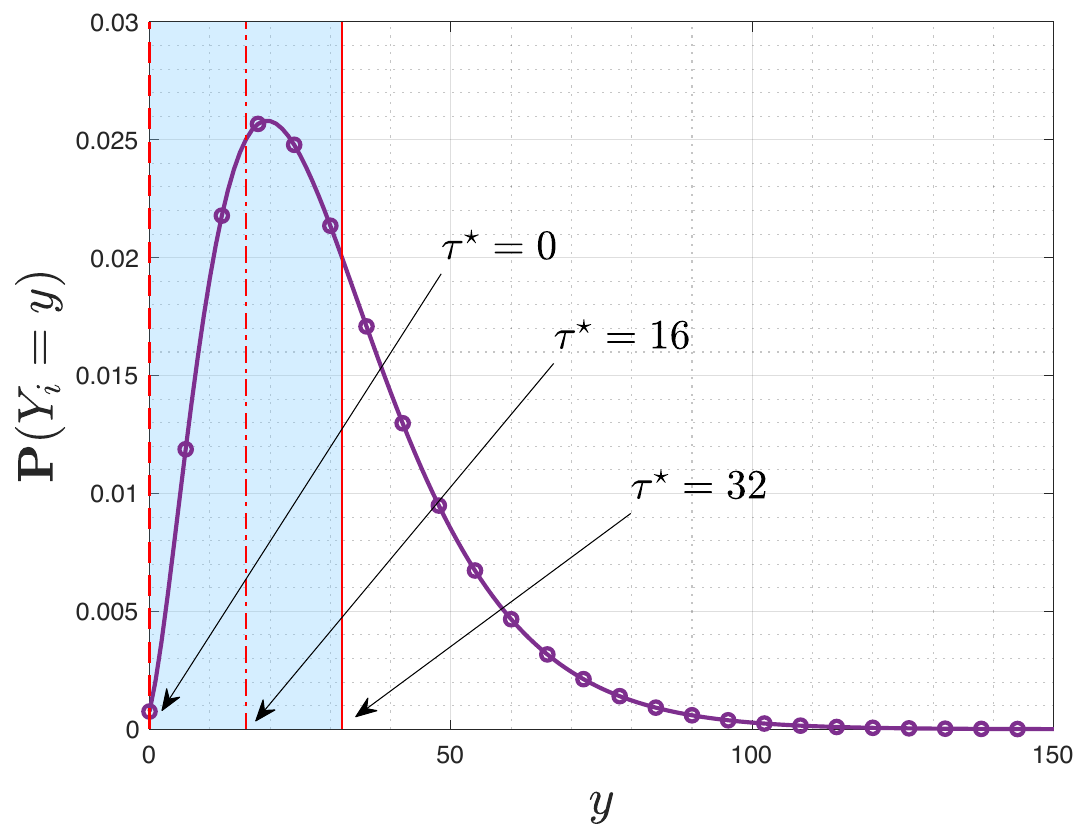}}
\hfill

\caption{Potential functions and an illustration of the optimal thresholds obtained from their maximization for different values of the parameters $k$, $\theta$ $\lambda$ and $g$ that specify the GPGG. For this set of numerical results, $p=1$, and therefore the BNE is of the low threshold type, $\mu^{\mathrm{low}}$.}
\label{fig:images}
    \end{figure*}

We consider three settings: (1) $p=k=1$, $\theta=1$, $\lambda=5$, for gains $g>0.3056$; (2) $p=1$, $k=1$, $\theta=0.5$, $\lambda=2$, for gains $g>2.592$; (3) $p=1$, $k=3$, $\theta=0.1$, $\lambda=1$, for gains $g>18.97$. Using the approach based on maximizing the potential function, we compute $\tau^\star$ by searching over the sets $\{0,1, \ldots, 50\}$, $\{0,1, \ldots, 75\}$, and $\{0,1, \ldots, 150\}$, respectively. The function $\Phi_{MF}(\tau)$ is displayed in \cref{fig:images} (a), (b) and (c). The resulting BNE thresholds $\tau^{\star}$ correspond to the unique maximizer of each of the unimodal functions displayed therein. The numerical results are reported in \cref{tab:example}, which also contains data about two alternative strategies: the omniscient threshold, $\tau_{\mathrm{omni}}$, and the certainty equivalence threshold, $\tau_{\mathrm{ce}}$. The omniscient threshold is computed by solving the equation:
\begin{equation}
\tau_{\mathrm{omni}} \Equaldef g^{1/p},
\end{equation}
as discussed in \cref{sec:omniscient}. The certainty equivalent threshold is obtained by using the mean-square estimate of the activation cost $\hat{c}_i(y)$ given in \cref{eq:MMSE} as if the agents where omniscient and $\hat{c}_i(y)$ were the true activation cost, resulting in the following implicit expression
\begin{equation}
 g = \frac{\Gamma(p+\tau_{\mathrm{ce}}+k)}{\Gamma(\tau_{\mathrm{ce}}+k)(\lambda+\theta)^p}.
\end{equation}
When $p=1$, the the certainty equivalence threshold $\tau_{\mathrm{ce}}$ can be evaluated in closed form as
\begin{equation}
\tau_{\mathrm{ce}}= (\lambda+\theta)g-k.
\end{equation}

\begin{table}
\centering
\caption{Thresholds obtained from optimizing the MFPF for different parameters defining the Gamma-Poisson Global Game with an infinite number of agents}
\begin{tabular}{|c|c|c|c|c|c|c|}
\hline
$\boldsymbol{k}$&$\boldsymbol{\theta}$ & $\boldsymbol{\lambda}$  & $\boldsymbol{g}$ & $\boldsymbol{\tau^\star}$ & $\boldsymbol{\tau_{\mathrm{omni}}}$ & $\boldsymbol{\tau_{\mathrm{ce}}}$ \\ \hline \hline
1&1 & 5 & 1 & 1 & 1 & 5 \\ 
1&1 & 5 & 2 & 5 & 2 & 11 \\ 
1&1 & 5 & 3 & 10 & 3 & 17 \\ \hline
2&0.5 & 2 & 5 & 3 & 5 & 10 \\ 
2&0.5 & 2 & 7.5 & 8 & 7.5 & 16 \\ 
2&0.5 & 2 & 10 & 13 & 10 & 23 \\ \hline
3&0.1 & 1 & 20 & 0 & 20 & 19 \\ 
3&0.1 & 1 & 40 & 16 & 40& 41 \\ 
3&0.1 & 1 & 60 & 32 &  60& 63 \\  \hline 
\end{tabular}

\label{tab:example}
\end{table}

\section{Modeling Bacterial Quorum Sensing Systems using Global Games}

\subsection{Fundamentals of Quorum Sensing}

\begin{figure}[ht!]
    \centering
    \includegraphics[width=0.8\columnwidth]{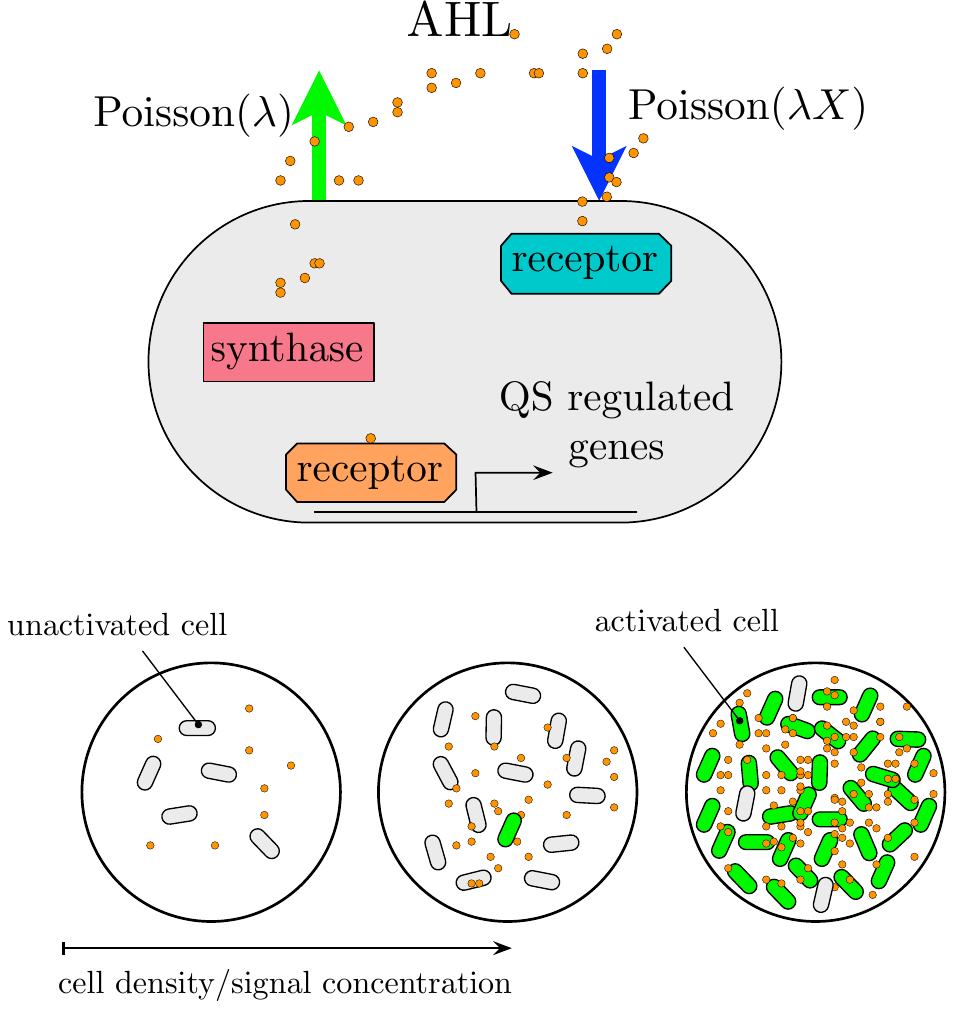}
    \caption{Signaling mechanism in quorum sensing: each bacterium emits and receives AHL molecules in and from the environment that are modeled according to Poisson random variables.}
    \label{fig:QS}
\end{figure}

Microbial consortia are distributed systems composed of multiple microorganisms that communicate via molecular signals to coordinate their collective behavior \cite{kylilis2018tools,martinelli2023multicellular}. A key element of such microbial coordination strategies is a process known as \textit{Quorum Sensing} (QS), a distributed decision-making mechanism used by bacteria to regulate density-dependent collective behavior.
In QS, bacteria release and sense molecules known as autoinducers. Each cell in the system releases autoinducers at a basal level, i.e., at a small constant rate. The production of autoinducers happens via enzymes called synthases. The type of synthase determines the class of autoinducer produced and is influenced by both the bacterial species and the gene being regulated \cite{bassler1999bacteria}. A cell may have several QS circuits, regulating many genes simultaneously. In this example, we will focus on Gram-negative bacteria controlling the expression of a single gene using the LuxI-LuxR mechanism, which is the simplest QS module found in nature. LuxI is an enzyme responsible for synthesizing \textit{acyl-homoserine lactones} (AHL), which is a type of autoinducer molecule. AHL molecules diffuse freely across the bacterial membrane into the surrounding environment and therefore are broadcast in the medium where the bacteria inhabit. \Cref{fig:QS} illustrates the LuxI-LuxR mechanism from the perspective of a single cell.

As the bacterial population grows, the concentration of AHL molecules in the environment increases. This accumulation depends on the density of the bacterial colony because more bacteria produce more AHL molecules. Therefore, the amount of autoinducers in the environment serves as a proxy for the colony density. The detection of AHL signal is performed by a transcriptional regulator protein (or receptor) called LuxR. 
When AHL reaches a threshold concentration, indicating high population density, they bind to LuxR in the cytoplasm. The binding stabilizes LuxR and activates it, forming an AHL-LuxR complex \cite{Boedicker:2015}.

The AHL-LuxR complex binds to specific DNA sequences called lux boxes in the promoter regions of target genes. This activates the transcription of these genes, leading to coordinated expression of group behaviors, such as: bioluminescence (in Vibrio fischeri),
virulence factor production (in pathogenic bacteria like Pseudomonas aeruginosa) and 
biofilm formation \cite{Gulec:2023}. The LuxI-LuxR module is also widely used in synthetic biology for the design of QS circuits to enable many bio-engineering applications \cite{Lindemann:2016,VanDerMeer:2010}.

\subsection{Mapping QS into a GGGP}

The purpose of QS is to promote the activation of genes depending on the density of the colony in the presence of partial state information. Although QS is a dynamical system, it evolves in a slow-varying time-scale, tipically measured in hours. Therefore, the density of the colony, which is the state variable, is assumed to be static over the relevant time horizon. The state is imperfectly observed by each bacterial agent. 

We begin by discussing the concept of density. Let $X$ represent the colony \textit{density}, defined as the number of bacteria per unit volume occupied by the colony. In our Bayesian model, the number of agents is assumed to be fixed, making the \textit{volume} an unknown random variable. Consequently, a larger volume corresponds to a lower density. It has been hypothesized that bacteria also use QS to jointly gain information about the local environment in addition to the colony density \cite{Ostovar:2020,west2012quorum}.

Density is always a positive quantity, and empirical studies of bacterial colonies in vivo often model the prior distribution of 
$X$ using the Gamma family of distributions \cite{GONZALESBARRON20111279}.

Prior to activating, every cell in the system emits auto-inducer \textit{signaling} molecules known as \textit{acyl homoserine lactone} (AHL) at a basal rate $\lambda>0$ \cite{Boedicker:2015,bassler1999bacteria,Michelusi:2017,Gangan:2022,Vasconcelos:2018b}. The emission of molecular signals is modeled by a Poisson variable of rate $\lambda$ \cite{Shaska:2023,Gulec:2023,michelusi2016queuing}. Since every cell in the colony produces AHL simultaneously, the number of signaling molecules aggregate in the environment. We assume that we have a closed system, and there is no signal diffusion. Therefore, if the density increases and the basal emission rate of autoinducers is fixed and equal to $\lambda$, the arrival of molecules at each cell is well-approximated by $Y_{i}\sim \mathcal{P}(\lambda X)$ \cite{Vasconcelos:2018}.

It is widely accepted that bacteria regulate the production of genes via \textit{threshold} policies. That means that once the concentration of AHL in the environment surpasses a certain level, the production of the appropriate gene is triggered. Threshold behaviors like this are ubiquitous in biological decision-making mechanisms~\cite{perkins2009strategies}. The fact that threshold policies are prevalent in QS indicate that they might be optimal in some sense. Indeed, by showing that threshold policies emerge as BNE of the GGGP we provide a theoretical evidence of the optimality of QS as a decision-making mechanism.

Finally, the local payoff function with a separable structure in the form of 
\begin{equation}
u_i(a_i,a_{-i},x) = a_i \cdot \Bigg( \underbrace{b\Big(\sum_{j\in [N]}a_j \Big)}_{\mathrm{public\ goods}}-\underbrace{c(x)}_{\mathrm{energetic\ cost}}\Bigg)
\end{equation}
is related to the so-called \textit{fitness} of the colony, and directly related to bacterial  growth rate when the QS process evolves over time \cite{darch2012density}. The fact that the activation cost is an increasing or decreasing function of the colony density is motivated from the following perspective: activating in a colony which has low density, may require much more energetic resources, reducing the growth rate (payoff) for the cell, which is the case for the bacterial strain \textit{Pseudomonas aeruginosa} using the LasI-LasR and RhlI-RhlR QS systems. In other cases, the energetic cost for activating at low densities may be low, and increase with the colony size, which is observed for \textit{Vibrio fischeri} LuxI-LuxR QS System.

We envision that this model can be used to synthetically design colonies with a desired threshold behavior. For instance, in localized drug-delivery systems when a certain enzyme is produced by a colony once a certain threshold concentration of bacteria is achieved. A separate application is on inferring parameters related to the payoff functions encoding the local activation preferences of natural systems, such as the the shape of the benefit function $b$ or the powerlaw coefficient $p$. A third application is the ability to quantify the activation accuracy of QS systems as a function of the ammount of biochemical noise in the environment. The aforementioned applications are left for future work.

\begin{remark}
It is interesting to note that the probability distribution of number of AHL molecules observed by the $i$-th bacterial agent in \cref{eq:observed_signal1} has three parameters ($\lambda$, ${\theta}$ and $k$), providing sufficient degrees of freedom to approximate empirical distributions in a wide range of applications. \Cref{fig:QS_Signal} illustrates the different shapes admitted by \cref{eq:observed_signal1} for different combination of parameters. Our model and the distribution for the number of signaling molecules observed by a cell coincides with the one independently obtained in \cite{Michelusi:2017}. 
\end{remark}

\section{Conclusions and Future work}

This paper introduces a framework for stochastic coordination games based on Global Games with a Gamma prior distribution and Poisson signals, which effectively captures the characteristics of molecular communications intrinsic to quorum sensing. We have established the existence of a Bayesian Nash Equilibrium within the class of threshold policies for any number of agents. The computation of ``optimal'' thresholds is a difficult problem for systems with a finite agents, which motivates the use a ``mean-field'' approach by computing the maximizer of a potential function in the limit of number of agents approaches infinity, a realistic assumption in the context of bioengineering applications. To the best of our knowledge, our model is the first to obtain structural results for non-Gaussian Global Games. 

The framework proposed herein is highly flexible and can be significantly generalized in several research directions. One such direction involves the decision to choose to activate one of multiple tasks (non-binary actions), rather than the single-task (binary actions) studied in this paper. Our results are based on the implicit assumption of a fully connected network, where there is no spatial distribution or arrangement among agents, i.e., every agent influences, and is influenced by, every other agent in the system. However, in practice, an agent typically influences only those in its immediate vicinity, leading to network effects. In the limit of a very large number of agents, the class of Graphon games \cite{Parise:2023} will play a critical role in this analysis. Finally, it remains an open question how to assess the coordination efficiency as a function of the hyperparameters that describe the stochastic environment of our system. In this regard, we are interested in establishing the fundamental limits of coordination for a given level of environmental noise using information-theoretic techniques.

\appendices

\section{Proof of Lemma 3}\label{sec:belief}

\begin{IEEEproof}
Let $i\neq j$. Consider the following conditional probability:
\begin{align}\label{eq:pi_low}
\mathbf{P}(Y_j=\ell & \mid Y_i=y)   =  \int_0^\infty \mathbf{P}(Y_j = \ell , X=x \mid Y_i=y)dx  \nonumber \\  & \stackrel{(a)}{=} 
\int_0^\infty \mathbf{P}(Y_j = \ell \mid X=x) f_{X\mid Y_i=y}(x)dx, 
\end{align}
where $(a)$ follows from the conditional independence of $Y_i$ and $Y_j$ given $X=x$. Since:
\begin{equation}\label{eq:conditional_poisson}
\mathbf{P}(Y_j = \ell\mid X=x) = \frac{(\lambda x)^\ell}{\ell!}e^{-\lambda x},
\end{equation}
the result follows by using \cref{lem:conditional_distribution} and \cref{eq:conditional_poisson} in \cref{eq:pi_low}.

\end{IEEEproof}

\section{Proof of Lemma 5}\label{sec:monotonicity}

\begin{IEEEproof}
Let us define the first-order difference
\begin{equation}
\Delta_{ij}^{\mathrm{low}}(y\mid \tau_j) \Equaldef \pi_{ij}^{\mathrm{low}}(y\mid \tau_j) - \pi_{ij}^{\mathrm{low}}(y+1\mid \tau_j).
\end{equation}
From the definition of $\pi^{\mathrm{low}}_{ij}$, we have
\begin{multline}\label{eq:Delta_Low}
\Delta_{ij}^{\mathrm{low}}(y\mid \tau_j) = \sum_{\ell=0}^{\tau_j}\left(\frac{\lambda}{2\lambda+\theta}\right)^\ell \left(\frac{\lambda +\theta}{2\lambda+\theta}\right)^{k+y} \times \\ \Bigg\{\binom{\ell+k+y-1}{\ell} - \binom{\ell+k+y}{\ell}\left(\frac{\lambda +\theta}{2\lambda+\theta}\right) \Bigg\}.
\end{multline}
From the Pascal Triangle identity, we have
\begin{equation}\label{eq:pascal_triangle}
\binom{\ell+k+y}{\ell} = \binom{\ell+k+y-1}{\ell -1} + \binom{\ell+k+y-1}{\ell},
\end{equation}
and after using \cref{eq:pascal_triangle} into \cref{eq:Delta_Low}, we obtain
\begin{multline}\label{eq:first_order}
\Delta_{ij}^{\mathrm{low}}(y\mid \tau_j) = \left(\frac{\lambda +\theta}{2\lambda+\theta}\right)^{k+y} \sum_{\ell=0}^{\tau_j}\Bigg\{\left(\frac{\lambda}{2\lambda+\theta}\right)^\ell \times \\ \binom{\ell+k+y}{k+y}   \left(\frac{(k+y-\ell)\lambda -\ell\theta}{(2\lambda+\theta)(\ell+k+y)} \right)\Bigg\}.
\end{multline}

The first step in the proof is to decompose the summation in \cref{eq:first_order} into two terms: the positive and the nonpositive part. Based on the inequality
\begin{equation}
(k+y-\ell)\lambda-\ell \theta>0 \Longleftrightarrow
\ell < \frac{k+y}{\lambda+\theta},
\end{equation}
we obtain \cref{eq:long_first_order}, in which the first term is  positive, whereas the second is nonpositive.
\begin{figure*}[!t]
\begin{multline}
\label{eq:long_first_order}
\Delta_{ij}^{\mathrm{low}}(y\mid \tau_j)  = \left(\frac{\lambda +\theta}{2\lambda+\theta}\right)^{k+y}  \Bigg[ \sum_{\ell=0}^{\min\{\lfloor \frac{k+y}{\lambda+\theta} \rfloor,\tau_j\}}\left(\frac{\lambda}{2\lambda+\theta}\right)^\ell  \binom{\ell+k+y}{k+y}   \Bigg(\frac{(k+y-\ell)\lambda -\ell\theta}{(2\lambda+\theta)(\ell+k+y)}\Bigg) \\ + \sum_{\min\{\lfloor \frac{k+y}{\lambda+\theta} \rfloor,\tau_j\}+1}^{\tau_j}\left(\frac{\lambda}{2\lambda+\theta}\right)^\ell  \binom{\ell+k+y}{k+y}   \left(\frac{(k+y-\ell)\lambda -\ell\theta}{(2\lambda+\theta)(\ell+k+y)}
\right) \Bigg].
\end{multline}
\centering --------------------------------------------------------------------------------------------------------------------------------------------------------

\end{figure*}

There are two possible regimes in \cref{eq:long_first_order} depending on the value of $\tau_j$.

\subsubsection*{Case 1}
\begin{equation}
\tau_j \leq \left\lfloor \frac{k+y}{\lambda +\theta} \right\rfloor \ \ \Longrightarrow \ \ \Delta_{ij}^{\mathrm{low}}(y\mid \tau_j)  >0.
\end{equation}

\subsubsection*{Case 2}
If $\tau_j$ satisfies
\begin{equation}
\tau_j \geq \left\lfloor \frac{k+y}{\lambda +\theta} \right\rfloor +1,
\end{equation}
then define the following function
\begin{multline}
A_{\tau_j}(y) \Equaldef \sum_{\left\lfloor \frac{k+y}{\lambda+\theta} \right\rfloor+1}^{\tau_j}\left(\frac{\lambda}{2\lambda+\theta}\right)^\ell  \times \\ \binom{\ell+k+y}{k+y}   \left(\frac{(k+y-\ell)\lambda -\ell\theta}{(2\lambda+\theta)(\ell+k+y)}
\right).
\end{multline}
In this regime, all the terms in \cref{eq:negative} are strictly negative. Therefore,
\begin{equation}\label{eq:bound_A_tau}
A_{\tau_j}(y) > A_{\infty}(y).
\end{equation}
From \cref{eq:bound_A_tau}, we obtain the following lower bound
\begin{multline}\label{eq:lower_bound_first_order}
\Delta_{ij}^{\mathrm{low}}(y\mid \tau_j)  > \left(\frac{\lambda +\theta}{2\lambda+\theta}\right)^{k+y}  \sum_{\ell=0}^{\infty}\left(\frac{\lambda}{2\lambda+\theta}\right)^\ell  \times \\ \binom{\ell+k+y}{k+y}   \Bigg(\frac{(k+y-\ell)\lambda -\ell\theta}{(2\lambda+\theta)(\ell+k+y)}\Bigg).
\end{multline}

The last step in the proof is to show that the right hand side of \cref{eq:lower_bound_first_order} is exactly equal to zero. We start by writing the identity below
\begin{equation}
\frac{(k+y-\ell)\lambda -\ell\theta}{(2\lambda+\theta)(\ell+k+y)} = \frac{\lambda}{2\lambda +\theta} - \left(\frac{\ell}{\ell +k+y} \right)
\end{equation}
From \cref{lem:identity}, we can obtain a combinatorial identity given by
\begin{equation}\label{eq:identity}
\sum_{\ell=0}^\infty \binom{\ell+k+y}{\ell} \Big(\frac{\lambda}{\theta+2\lambda} \Big)^\ell = \Big(\frac{\theta+\lambda}{\theta+2\lambda} \Big)^{-(k+y)}.
\end{equation}

Using \cref{eq:identity} in \cref{eq:lower_bound_first_order}, we have
\begin{multline}
\Delta_{ij}^{\mathrm{low}}(y\mid \tau_j)  > \frac{\lambda}{2\lambda +\theta} 
- \textbf{}\left(\frac{\lambda +\theta}{2\lambda+\theta}\right)^{k+y} \times \\  \underbrace{\sum_{\ell=0}^{\infty} \binom{\ell+k+y}{k+y}\left(\frac{\ell}{\ell+k+y} \right)\left(\frac{\lambda}{2\lambda+\theta}\right)^\ell}_{\Equaldef \circledast} .   
\end{multline}

Then,
\begin{equation}
\circledast =  \sum_{\ell=1}^{\infty}  \binom{\ell+k+y-1}{\ell-1}   \left(\frac{\lambda}{2\lambda+\theta}\right)^\ell \stackrel{(a)}{=}\left(\frac{\lambda +\theta}{2\lambda+\theta}\right)^{-(k+y)},
\end{equation}
where $(a)$ follows from the change of variables $\ell'=\ell-1$, and using the identity in \cref{eq:identity}. Therefore, $\Delta_{ij}^{\mathrm{low}}(y\mid \tau_j)  > 0.$

Since the first order difference $\Delta_{ij}^{\mathrm{low}}(y\mid \tau_j)$ is strictly positive for all $y$, it implies that $\pi^{\mathrm{low}}_{ij}(y \mid \tau_j)$ is strictly monotone decreasing. Moreover, $\pi^{\mathrm{low}}_{ij}(y \mid \tau_j) \rightarrow 0$. Consequently, from the relationship between $\pi^{\mathrm{low}}_{ij}(y \mid \tau_j)$ and $\pi^{\mathrm{high}}_{ij}(y \mid \tau_j)$, we have
\begin{equation}
\pi^{\mathrm{high}}_{ij}(y+1 \mid \tau_j) - \pi^{\mathrm{high}}_{ij}(y\mid \tau_j) = \Delta_{ij}^{\mathrm{low}}(y\mid \tau_j) > 0,
\end{equation}
which implies that $\pi^{\mathrm{high}}_{ij}(y\mid \tau_j)$ is strictly monotone increasing. Moreover, $\pi^{\mathrm{high}}_{ij}(y \mid \tau_j) \rightarrow 1$.

\end{IEEEproof}

\vspace{-0.25in}

\bibliographystyle{IEEEtran}
\bibliography{QSGG.bib}

\begin{IEEEbiographynophoto}
{Marcos M. Vasconcelos} is an Assistant Professor with the Department of Electrical Engineering at the FAMU-FSU College of Engineering, Florida State University. He received his Ph.D. from the University of Maryland, College Park, in 2016. He was a Research Assistant Professor at the Commonwealth Cyber Initiative and the Bradley Department of Electrical and Computer Engineering at Virginia Tech from 2021 to 2022. From 2016 to 2020, he was a Postdoctoral Research Associate in the Ming Hsieh Department of Electrical Engineering at the University of Southern California. His research interests include networked control and estimation, robotic networks, game theory, distributed optimization, distributed learning, and systems biology.
\end{IEEEbiographynophoto}
\vspace{-0.25in}

\begin{IEEEbiographynophoto}{Behrouz Touri} an Associate Professor of Industrial and Systems Engineering at the University of Illinois at Urbana-Champaign. He earned his B.Sc. in Electrical Engineering from Isfahan University of Technology, Iran (2006), his M.Sc. in Communications, Systems, and Electronics from Jacobs University, Germany (2008), and his Ph.D. in Industrial Engineering from the University of Illinois Urbana-Champaign (2011).

He served as an Assistant Professor of Electrical Engineering at the University of Colorado Boulder (2014-2017) and was a faculty member with the ECE Department at the University of California San Diego (2017-2025). His research interests include applied probability theory, distributed optimization, control and estimation, population dynamics, and evolutionary game theory. He was awarded the American Control Council’s Donald P. Eckman Award in 2018.
\end{IEEEbiographynophoto}

\end{document}